\documentclass[onecolumn,10pt]{aastex631}

\usepackage{amsmath,bm,mathrsfs}
\usepackage{graphicx,color}
\usepackage{natbib}
\usepackage{amsbsy}
\DeclareFontFamily{OMS}{bcmsy}{}
\DeclareFontShape{OMS}{bcmsy}{m}{n}{%
	<5> bcmsy5
	<6> bcmsy6
	<7> bcmsy7
	<8> bcmsy8
	<9> bcmsy9
	<10-> bcmsy10}{}
\DeclareSymbolFont{symbols}{OMS}{bcmsy}{m}{n}

\newcommand{\imag}{\mathrm{i}}
\newcommand{\uk}{\bm{\hat k}}
\newcommand{\vk}{\bm{k}}
\newcommand{\ek}[1]{\bm{e}_#1(\uk)}

\newcommand{\<}{{\kern-5pt}}

\newcommand{\thrj}[6]{\biggl( %
	\arraycolsep .25em
	\begin{matrix}
	#1 &#2 &#3 \\
	#4 &#5 &#6
	\end{matrix}\biggr)}
\newcommand{\sixj}[6]{\biggl\{
	\arraycolsep .25em
	\begin{matrix}
	#1&#2&#3\\
	#4&#5&#6\\
	\end{matrix}\biggr\}}

\def\shalf{^{\!\frac{1}{2}}}
\def\mlangle{\kern.175em\langle}	
\def\mrangle{\rangle\kern.175em}	
\def\ket#1{|{#1}\mrangle}
\def\bra#1{\mlangle{#1}|}
\def\pro#1{\ket{#1}\bra{#1}}
\def\redmat#1#2#3{\bra{#1}\kern -1pt|#2|\kern -1pt\ket{#3}}

\newcommand{\apx}[1]{^{\mbox{\tiny{(#1)}}}}

\begin{document}

\title{%
A unifying polarization formalism 
for electric- and magnetic-multipole interactions}

\author{%
R.\ Casini}
\affiliation{%
High Altitude Observatory, National Center for Atmospheric
Research, P.O.\ Box 3000, Boulder, CO 80307-3000, U.S.A.}
\author{%
R.\ Manso Sainz}
\affiliation{%
Third Institute of Physics, University of G\"ottingen, 
Friedrich-Hund-Platz 1,
37077 G\"ottingen, Germany}
\author{%
A.\ L\'opez Ariste}
\author{%
N.\ Kaikati}
\affiliation{%
IRAP, Universit\'e de Toulouse, CNRS, CNES, UPS.\  
14, Av.\ E.\ Belin. 31400 Toulouse, France}

\begin{abstract}
We extend the spherical tensorial formalism for polarization to the
treatment of electric- and magnetic-multipole transitions of any order. 
We rely on the spherical-wave expansion to derive the tensor form of 
the operator describing the interaction of the atomic system with a 
polarized radiation field, which naturally leads to the introduction 
of spherical tensors describing the polarization properties of the
interacting field. As a direct application, the formalism is used 
to model the radiation anisotropy affecting the scattering of radiation 
in an electric-quadrupole transition, and the associated Hanle effect 
in the presence of a magnetic field.
\end{abstract}



\defcitealias{LL04}{LL04} 
\defcitealias{SM68}{SM68}
\defcitealias{VMK88}{VMK88}

\section{Introduction} \label{sec:intro}

The study of the polarization of the electromagnetic radiation we receive
from astrophysical objects is at the basis of any diagnostics of
non-isotropic physical processes in the emitting gas. Typical examples
are the presence of deterministic magnetic and electric fields that
break the symmetry of the atom-photon interaction process, and non-isotropic
excitation -- both radiative and collisional -- of plasma, which can
lead to significant departures of the plasma conditions from local
thermodynamic equilibrium (LTE).
The importance of non-LTE conditions in 
astrophysical plasmas for the production and transfer of polarized 
radiation in spectral lines has long been recognized. For example,
a comprehensive understanding of non-LTE effects is critical for our 
ability to model polarized signals in the solar spectrum, and how
these are modified by the presence of magnetic and electric fields in 
the solar atmosphere. This is at the basis of most solar magnetism 
diagnostics \cite[e.g.,][]{CL08}.

The largest majority of spectral lines from neutral and ionized atomic
species observed in the solar atmosphere and other astrophysical plasmas 
are permitted transitions, which originate in the interaction of the 
radiation field with the induced electric dipole of the optical electron 
in the ion. For this reason, the modeling of these lines has received 
most of the attention of the heliophysics community since the very
inception of solar magnetism investigations \citep[e.g.,][]{Ha08}. As
these studies even predate the development of quantum mechanics,
the modeling of polarized line formation was originally based on 
Lorentz's classical theory of the bound optical electron as a damped, 
three-dimensional oscillator forced by the radiation field \citep{Lo52}, 
and how the spectral shape and polarization of the radiated energy is 
affected by the presence of a magnetic field via the Zeeman effect. 
The complexity of atomic spectra observed in both laboratory gases and 
astrophysical plasmas was largely responsible for the development of 
the quantum mechanical description of atomic transitions
\cite[e.g.,][]{So23}, 
and how magnetic and electric fields modify the absorbed and re-emitted 
radiation \cite[see, e.g.,][]{BS57}. 

However, along with the prevalent permitted lines, forbidden transitions are 
also often observed in astrophysical contexts. In fact, they were discovered 
in the spectra of planetary nebulae, the solar corona, and also in the Earth's 
aurorae, even before than in laboratory plasmas
\citep{Bo36,Pa40,Ga68,Ei81}. This was possible because, in the typical 
rarefied state of those environments, the upper levels of transitions with 
relatively low oscillator strengths can often survive long enough for 
spontaneous radiative decay to occur. The ensuing lines are thus optically 
thin, and play 
an important role for diagnosing chemical abundances, temperature, density, 
and velocity fields in various astrophysical contexts, including the Sun
\citep{Do75,Sa77,Ch79}, young stars \citep{Ap84,Ha94,CR97}, stellar
winds \citep{Sa93,Si16}, nebulae \citep{Se53,Ts03}, and active
galactic nuclei \cite[AGN;][]{Pe85}.  In environments with intense nearby 
radiation sources, such as the solar corona, gaseous nebulae, and AGNs, the 
incoming radiation may be anisotropically
scattered, producing distinctive line polarization signatures that can probe 
the geometry 
of unresolved plasma structures, large-scale mass motions, and magnetic fields 
via the Hanle effect \citep{Le94,Ju98,Ju01,Ca17}. As an example, magnetic-dipole 
emission lines from highly ionized atomic species are regularly observed in 
the solar corona \citep{Ju98}, and commonly used for magnetism diagnostics 
\citep{Ch65,Ho74,SB77,CJ99,Li17}. However, the contribution of electric 
quadrupole and higher multipoles, although less significant, can be important 
in certain cases of astrophysical interest \citep{Se55,Ga57,Ga62a,Qu96}. Some 
electric-quadrupole lines have also been predicted in the relatively denser 
environment of the solar photosphere \citep{Ga62}, although detection
conditions are exceedingly difficult in that case. 
Another notable example is represented by the transitions among the
lowest energy levels of \ion{O}{1}, \ion{O}{2}, and \ion{O}{3}, which are 
exclusively of the magnetic dipole and electric quadrupole types. These 
atomic species are important tracers of the dynamics of \ion{H}{2} 
star-forming regions \citep[e.g.,][]{BGH1974,Wilsonetal2024} 
and have formerly been used \citep{LLB1987} or suggested 
\citep{YL2006,YL2008,YL2012} as valuable tools for the determination 
of magnetic fields in these regions. However the importance of the 
contribution of the E2 transitions of those systems to the final 
observed transition has escaped attention. H$\alpha$ linear 
spectro-polarimetry has been a common diagnostic tool in Herbig Ae/Be 
and T Tauri stars \citep{VinkETAL2002,VinkETAL2005}. The very same atomic 
lines of the \ion{O}{1}, \ion{O}{2} and \ion{O}{3} systems are 
visible in these objects, although linear polarimetry of these lines 
is scarce \citep{AOV2016}. Future polarimetry of 
these objects, often associated with the presence of disks, 
requires a correct understanding of the polarization of E2 lines. 
In order to exploit the diagnostic potential of these
forbidden transitions, a unified theoretical approach becomes necessary
to model self-consistently the various multipolar contributions
to the scattered radiation.

The application of quantum-mechanical methods to the description of
polarized line formation in permitted transitions is nowadays well established
\citep{LTH71,La83,St94,LL04}.
In the context of astrophysical plasmas, the complex non-locality of the
radiation transfer problem in an optically thick medium represents a
difficult numerical challenge. This is the well-known problem of
non-LTE, which in its most benign form leads to the departure of the atomic 
level populations from the Maxwell-Boltzmann statistical distribution in
response to the out-of-equilibrium, ambient radiation field, but 
which in its most general form also leads to the manifestation of quantum 
interference and population imbalances among the magnetic levels,
producing a complex variety of polarization signatures. While 
in the first case a description of the magnetic effects with the 
interacting atom-photon system can be limited to the Zeeman and the (incomplete) 
Paschen-Back effects, the most general case -- often dubbed the ``non-LTE problem 
of the 2nd kind'' \citep{La83} -- fully manifests the quantum-mechanical nature of
the problem, requiring a complete grasp of the density matrix description of 
the partially polarized statistical ensemble of interacting atoms and photons.

The standard formalism that describes this problem in its generality,
with applications to the study of astrophysical plasmas, is exhaustively 
presented in the monograph by \citet[hereafter \citetalias{LL04}]{LL04}, 
with the notable physical restriction that the frequency of the incident 
photon is assumed to be completely redistributed (CRD hypothesis) during 
the interaction process. Through this assumption, the scattering of 
radiation can be described as the incoherent succession of single-photon 
absorption and re-emission \cite[e.g.,][]{Sa67}. 
Such a formalism relies on the framework of the irreducible spherical 
tensors \citep{Fa57}, which allows one to describe the process of
atom-photon interaction in the form of a multipolar expansion of 
products of tensorial components of the atomic density 
matrix---accounting for the excitation state of the atomic 
system---and the radiation field---describing the geometric distribution 
and polarization of the
incident and scattered radiation. Such an approach was ushered into 
the field of magnetism diagnostics of astrophysical objects in the 
1970's and early 1980's \citep{LTH71,SB77,Bo80,La83}, and it has now
become mainstream in solar physics. This formalism provides a very 
elegant, compact, and physically insightful 
description of the polarized scattering of radiation by a magnetized plasma. 
\citetalias{LL04} gives a comprehensive description of the 
formalism and of many astrophysical applications, but it is 
predominantly focused on permitted, electric-dipole 
transitions.\footnote{Because of 
the applicability of the same tensorial formalism to the case of 
magnetic-dipole transitions, \citetalias{LL04} also briefly
consider the line formation problem of the corresponding forbidden
transitions (see \S6.8 of that monograph).}


In order to generalize the current CRD theory of polarized line formation 
to the treatment of electric and magnetic multipole radiative transitions 
of order $l$ (for brevity, hereafter we refer to these simply 
as E$l$ and M$l$ transitions), 
it is necessary to extend such a tensorial formalism to atomic transitions 
of any multipolar order. In order to do so, we first summarize the method 
of spherical-wave expansion of the interaction Hamiltonian. While this 
derivation is not new, exhaustive presentations of the problem are 
scarce in the literature, and they consistently rely on personal choices 
of notation that make the comparison of results quite difficult. We hope 
to provide here a unifying formalism that is going to be more accessible 
to scientists working in the field of solar and stellar spectro-polarimetry, 
being based on the standard notation and formalism of \citetalias{LL04}. 
We show how this expansion naturally leads to the introduction of new 
generalized spherical tensors (in both reducible and irreducible forms) 
necessary to model the polarized line formation of multipolar transitions. 
We also rely on the monograph by \citet[hereafter \citetalias{VMK88}]{VMK88} 
for many definitions, and on the exhaustive monograph by
\citet[hereafter \citetalias{SM68}]{SM68} for 
the justification of several key approximations.

From the outset, we will limit our study to the so-called \emph{long-wavelength 
regime}, $2\pi\,a_0/\lambda\ll 1$, where $a_0$ is Bohr's radius. This enables 
applications of the presented formalism for $\lambda\gtrsim 10$\,nm, spanning 
most of the spectral range of interest for the magnetic diagnostics of
solar and astrophysical plasmas.

\section{General formalism} \label{sec:genform}

\begin{figure}
\centering
\includegraphics[height=.51\textheight]{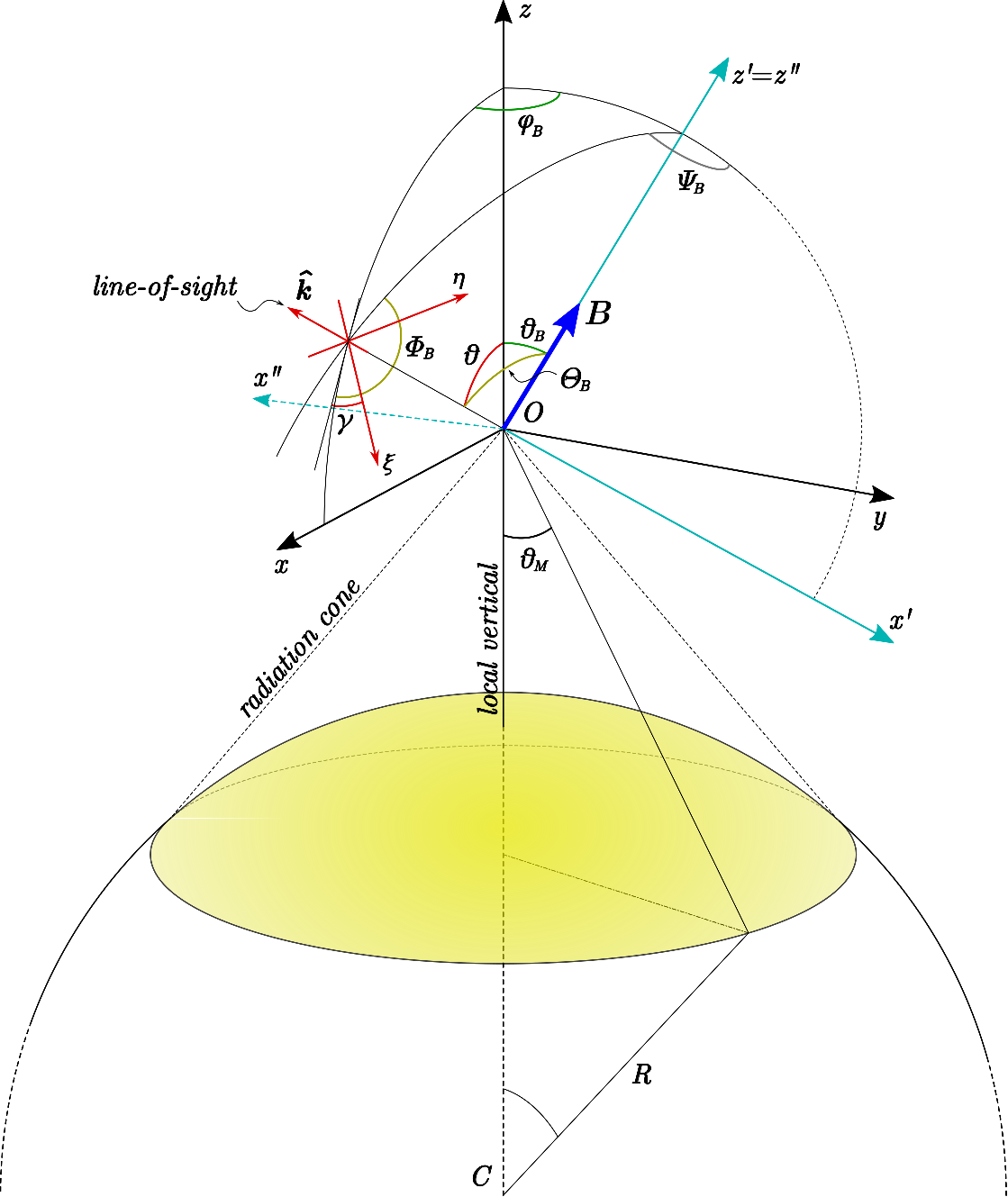}
\caption{\label{fig:scatgeom}
Geometry describing radiation scattering in the presence of a 
magnetic field. All the relevant angles for the transformation between
the solar frame $S\equiv(O;xyz)$ and the magnetic frame
$B\equiv(O;x'y'z')$ are shown. The scattered radiation 
field is specified in the observer's frame $(\xi,\eta,\uk)$,
where $\xi$ identifies the reference direction for linear 
polarization. We note that this geometry corresponds to the choice 
$\varphi=0$ for the triplet $(\vartheta,\varphi,\gamma)$ needed to 
specify the polarization tensors ${\cal T}^{l:K}_Q(i,\uk)_S$ in the solar 
frame (cf.~Table~\ref{tab:geom_tens}), as the line-of-sight
lies in the $xz$-plane of that reference frame.}
\end{figure}

The entire formalism for the modeling of 
E$l$ and M$l$
transitions is based on the derivation of the matrix element 
of the interaction Hamiltonian of the atom with a polarized radiation field. 
One can start from the expression of the interaction Hamiltonian operator 
given by \citetalias{LL04} (cf.~eq.~[6.18]) between an electron of
momentum $\bm{p}$ and a generally polarized 
transverse field of angular frequency $\omega$, propagation vector 
$\vk=(\omega/c)\,\uk$, and (circular) polarization state 
$\beta=\pm 1$ on the plane normal to $\uk\equiv(\vartheta,\varphi)$ 
(see Fig.~\ref{fig:scatgeom}),
\begin{equation} \label{eq:HintQ}
Q(\omega,\uk,\beta)
=d_\omega\sum_i
	\exp(\imag\,\vk\cdot\bm{r}_i)\,\ek{\beta}\cdot\bm{p}_i\;,
\qquad (\beta=\pm 1)
\end{equation}
where, using standard notation (we assume $e>0$ throughout),
\begin{equation} \label{eq:d_coeff}
d_\omega\equiv\frac{e}{m_{\rm e}}
	\biggl(\frac{2\pi\hbar}{\omega {\cal V}}\biggr)\shalf
	\equiv\frac{e}{m_{\rm e}}\,c_\omega\;.
\end{equation}
Here, $\cal V$ is the volume of the quantization box, which ultimately
is replaced by the volumetric density of the emitter.

It is important to remark that the expression (\ref{eq:HintQ}) only
considers the $\bm{A}\cdot\bm{p}$ part of the photon-atom
interaction Hamiltonian. This is completely adequate to treat 
E$l$ radiative transitions in the long-wavelength
approximation (see discussions in \citetalias{LL04}, after eq.~[6.15],
and in \citetalias{SM68}, after eq.~[10.(7.10)]). On
the other hand, in treating 
M$l$
transitions, the
interaction of the electron spin $\bm{S}$ with the magnetic component 
of the radiation field is comparable to that of the electron's 
orbital angular momentum (similarly to the case of the Zeeman effect 
in a static magnetic field), and the interaction Hamiltonian must take 
into account such additional contribution, which has the form
$\hbar\bm{S}\cdot\mathbf{curl}\,\bm{A}$ (cf.~\citetalias{LL04}, 
eq.~[6.15]; also \citetalias{SM68}, eq.~[10.(4.16)]).

In order to express the interaction Hamiltonian as a series of
multipole orders beyond the electric-dipole approximation, it is 
advantageous to exploit a spherical-wave expansion of the phase 
factor $\exp(\imag\vk\cdot\bm{r})$ in the expression 
(\ref{eq:HintQ}), as this leads to a unified formalism
for treating both electric and magnetic multipoles of any order. 
This derivation is given in details in App.~\ref{sec:spherwave}. 
On the other hand, if one is only interested in deriving the 
magnetic-dipole 
(M1) and electric-quadrupole (E2)
contributions, a 
physically more insightful approach consists in decomposing the dyadic 
product $(\vk\cdot\bm{r})(\ek{\beta}\cdot\bm{p}$), arising from the
first-order Taylor expansion of the phase factor, as the sum of two 
contributions, a symmetric one, corresponding to the electric-quadrupole
term, and an anti-symmetric one, corresponding to the magnetic-dipole
term (e.g., \citealt{Sa67}, \citetalias{LL04}). Such a decomposition is 
readily achieved using the Binet-Cauchy identity, and it is
presented in App.~\ref{sec:dyadic} for explicitly deriving the 
electric-quadrupole term. The same decomposition approach, while in 
principle still possible, becomes exceedingly cumbersome and less intuitive
for deriving the subsequent terms of the multipolar expansion of the 
interaction Hamiltonian. Hence, the need to rely on a unified approach,
which also naturally leads to the development of the tensorial
formalism necessary to treat polarized line formation for multipolar
transitions.
%

The main result of the spherical-wave expansion of the interaction
Hamiltonian is represented by the following matrix elements 
for the electric and magnetic contributions to
the interaction operator (\ref{eq:HintQ}) describing the 
radiative transition
between two atomic states $\ket{m}$ and $\ket{n}$, in the presence of a
monochromatic radiation field of propagation vector $\vk$, and state of 
polarization $\beta=\pm1$
(see App.~\ref{sec:spherwave}, 
eqs.~[\ref{eq:qmat1}] and [\ref{eq:mmat1}]),
\begin{subequations}
\begin{eqnarray} 
\label{eq:qmat1.summary}
\bigl[Q(\omega,\uk,\beta)_\mathrm{E}\bigr]_{mn}
&\simeq&-\imag\,c_\omega\,\omega_{mn} \sum_{K}
	\frac{\imag^{K-1}\,k^{K-1}}{(2K-1)!!}
	\left(\frac{K+1}{2K}\right)\shalf
	\sum_{Q}(-1)^{Q}\,(\mathscr{Q}^K_{-Q})_{mn}\,D^K_{\beta Q}(\hat R)\;, \\
\label{eq:mmat1.summary}
\bigl[Q(\omega,\uk,\beta)_\mathrm{M}\bigr]_{mn}
&\simeq&c_\omega\,\omega_{mn}\,\beta \sum_K
	\frac{\imag^{K-1} k^{K-1}}{(2K-1)!!}
	\left(\frac{K+1}{2K}\right)\shalf
	\sum_Q (-1)^{Q} (\mathscr{M}^K_{-Q})_{mn}\,D^K_{\beta Q}(\hat R)\;,
\end{eqnarray}
\end{subequations}
where 
$\hat{R}=\hat{R}_z(-\gamma)\hat{R}_y(-\vartheta)\hat{R}_z(-\varphi) 
\equiv(-\gamma,-\vartheta,-\varphi)$ is the rotation from the observer's 
frame to the solar frame (see Fig.~\ref{fig:scatgeom}), and
$\mathscr{Q}^K_Q$ and $\mathscr{M}^K_Q$ are the electric- and magnetic-multipole 
components of order $K$, expressed in terms of spherical
harmonics $Y^K_Q$
(see App.~\ref{sec:spherwave}, eqs.~[\ref{eq:Emultipole}] and
[\ref{eq:Mmultipole}]; cf.~also \citetalias{SM68}, eqs.~[10.(7.11)] 
and [10.(6.8$a$)]),
\begin{subequations}
\begin{eqnarray} 
\label{eq:Emultipole.Q}
\mathscr{Q}^K_Q
&=&-\biggl(\frac{4\pi}{2K+1}\biggr)\shalf
	e \sum_i r_i^K Y^K_Q(\bm{\hat r}_i)\;, \\
\label{eq:Mmultipole.Q}
\mathscr{M}^K_Q
&=&-\biggl(\frac{4\pi}{2K+1}\biggr)\shalf \mu_0
	\sum_i \mathbf{grad}_i\bigl\{r_i^K Y^K_Q(\bm{\hat r}_i)\bigr\}
	\cdot\left(\frac{2\bm{L}_i}{K+1}+2\bm{S}_i\right)\;,
\end{eqnarray}
\end{subequations}
where $\mu_0=e\hbar/(2m_{\rm e}c)$ is Bohr's magneton. We note how
the expressions (\ref{eq:qmat1.summary}) and (\ref{eq:mmat1.summary}) 
are formally identical, the only difference being the presence of an additional 
factor $\imag\beta$ in the magnetic contribution.

It is instructive to derive the expressions of these
multipole operators to the lowest orders $l=1,2$.
For the electric-multipole case, we immediately find from
eq.~(\ref{eq:Emultipole.Q}) the well-known expressions
\begin{subequations}
\begin{eqnarray} \label{eq:EM1}
\mathscr{Q}^1_Q
&=&-e\sum_i r_i\,(\bm{\hat r}_i)_Q=-e\,r_Q \\
\mathscr{Q}^2_Q
&=&-\left(\frac{4\pi}{5}\right)\shalf e 
\sum_i r_i^2 Y^2_Q(\bm{\hat r}_i)\;.
\end{eqnarray}
\end{subequations}
For calculating the magnetic-multipole case, we can use 
eq.~[5.8.3.(9)] of \citetalias{VMK88} to evaluate the 
$\mathbf{grad}_i\{\,\}$ factors in eq.~(\ref{eq:Mmultipole.Q}),
\begin{displaymath}
\mathbf{grad}\bigl\{r^K Y^K_Q(\bm{\hat r})\bigr\}
=\left[K(2K+1)\right]\shalf r^{K-1}\,\bm{Y}^{K-1}_{KQ}(\bm{\hat r})\;,
\end{displaymath}
where $\bm{Y}^l_{KQ}$ are vector spherical harmonics (see
App.~\ref{sec:spherwave}, eq.~[\ref{eq:VSH}] for the definition). 
We thus find,
\begin{subequations}
\begin{eqnarray} \label{eq:M1}
\mathscr{M}^1_Q
&=&-(4\pi)\shalf \mu_0 \sum_i 
	\bm{Y}^0_{1Q}(\bm{\hat r}_i)\cdot(\bm{L}_i+2\bm{S}_i) 
=-\mu_0 \sum_i 
	\bm{e}^\ast_Q\cdot(\bm{L}_i+2\bm{S}_i) \nonumber \\
&=&-\mu_0\,
	(\bm{L}+2\bm{S})_Q\;, \\
\label{eq:M2}
\mathscr{M}^2_Q
&=&-\left(\frac{4\pi}{5}\right)\shalf \mu_0 \sum_i 
	\sqrt{10}\,r_i\,
	\bm{Y}^1_{2Q}(\bm{\hat r}_i)\cdot
	\left(\frac{2}{3}\,\bm{L}_i+2\bm{S}_i\right) 
=-(8\pi)\shalf \mu_0 \sum_i 
	r_i\,\sum_{pq} (1p,1q|2Q)\,Y^1_p(\bm{\hat r}_i)
	\left(\frac{2}{3}\,\bm{L}_i+2\bm{S}_i\right)_{\!\!q}
	 \nonumber \\
&=&-\left(\frac{8}{3}\right)\shalf \mu_0 \sum_i 
	r_i \bigl[\left(\frac{4\pi}{3}\right)\shalf
	\bm{Y}^1(\bm{\hat r}_i)\otimes
	(\bm{L}_i+3\bm{S}_i)\bigr]^2_Q
\end{eqnarray}
\end{subequations}
(cf.\ also \citetalias{SM68}, eq.~[10.(6.9$b$)]).

\subsection{Spherical tensors for multipolar interactions}
\label{sec:tensors}

The expressions (\ref{eq:qmat1.summary}) and (\ref{eq:mmat1.summary}) 
show that the calculation of the matrix element of the
multipolar operator $\mathscr{O}^l$
of order $l$, between two atomic states $\ket{a}$ and $\ket{b}$,
generally implies the evaluation of products of the forms,
\begin{displaymath}
\sum_q (-1)^q\,(\mathscr{O}^l_{-q})_{ab}\,D^l_{\beta q}(\hat{R})\;,\qquad
\imag\beta\sum_q (-1)^q\,(\mathscr{O}^l_{-q})_{ab}\,D^l_{\beta q}(\hat{R})\;.
\end{displaymath}
The calculation of the corresponding atomic transition probabilities 
thus involves products of the forms
\begin{eqnarray*}
\sum_{pq} (-1)^{p+q}\,(\mathscr{O}^l_{-p})_{ab}\,(\mathscr{O}^l_{-q})_{cd}^\ast\,
	D^l_{\alpha p}(\hat{R})\,D^l_{\beta q}(\hat{R})^\ast
&=&\sum_{pq} (-1)^{p+q}\,(\mathscr{O}^l_{-p})_{ab}\,(\mathscr{O}^l_{-q})_{cd}^\ast\;
	{\cal E}^l_{pq}(\alpha,\beta;\uk)_E\;, \\
\alpha\beta\sum_{pq} (-1)^{p+q}\,(\mathscr{O}^l_{-p})_{ab}\,(\mathscr{O}^l_{-q})_{cd}^\ast\,
	D^l_{\alpha p}(\hat{R})\,D^l_{\beta q}(\hat{R})^\ast
&=&\sum_{pq} (-1)^{p+q}\,(\mathscr{O}^l_{-p})_{ab}\,(\mathscr{O}^l_{-q})_{cd}^\ast\;
	{\cal E}^l_{pq}(\alpha,\beta;\uk)_M\;,
\end{eqnarray*}
where we introduced the dyadic tensors for the electric and magnetic 
multipolar cases (cf.~\citetalias{LL04}, eq.~[5.142]),
\begin{subequations}
\label{eq:Edef}
\begin{eqnarray} 
{\cal E}^l_{pq}(\alpha,\beta;\uk)_\mathrm{E}
&=& D^l_{\alpha p}(\hat{R})\,D^l_{\beta q}(\hat{R})^\ast\;, \\
{\cal E}^l_{pq}(\alpha,\beta;\uk)_\mathrm{M}
&=& \alpha\beta\,D^l_{\alpha p}(\hat{R})\,D^l_{\beta q}(\hat{R})^\ast\;.
\end{eqnarray}
\end{subequations}
Since $\alpha,\beta=\pm 1$, evidently
\begin{displaymath}
{\cal E}^l_{pq}(\beta,\beta;\uk)_\mathrm{E}
\equiv {\cal E}^l_{pq}(\beta,\beta;\uk)_\mathrm{M}\;.
\end{displaymath}
These dyadic tensors satisfy conjugation and trace properties similar to those
given by \citetalias{LL04} for the dipole case $l=1$.
In particular, the conjugation relation eq.~(5.145) remains identical,
as well as the trace over the spherical-basis components ($p,q$),
eq.~(5.144), since
\begin{equation} \label{eq:Etens_contr0}
\sum_q\,
{\cal E}^l_{qq}(\alpha,\beta;\uk)_\mathrm{M}=\alpha\beta\,\delta_\alpha^\beta
=\alpha^2\,\delta_\alpha^\beta=\delta_\alpha^\beta=
\sum_q\,
{\cal E}^l_{qq}(\alpha,\beta;\uk)_\mathrm{E}\;.
\end{equation}
Instead, the trace over the radiation modes, eq.~(5.143), 
generalizes to the following
\begin{equation} \label{eq:Etens_contraction}
\sum_{\beta=\pm 1}\oint\frac{d\uk}{4\pi}\;
{\cal E}^l_{pq}(\beta,\beta;\uk)_\mathrm{E,M}=\frac{2}{2l+1}\,\delta_p^q\;.
\end{equation}
Both sum rules (\ref{eq:Etens_contr0}) and (\ref{eq:Etens_contraction})
follow directly from the orthogonality properties of the rotation matrices 
(e.g., \citetalias{VMK88}, eq.~[4.1.(6)]).
Analogously to eq.~(5.146) of \citetalias{LL04}, we can 
introduce a reducible set of polarization tensors
\begin{equation} \label{eq:Ttens}
{\cal T}^l_{pq}(i,\uk)_\mathrm{E,M}=\frac{1}{2}\sum_{\alpha\beta=\pm1}
	(\bm{\sigma}_i)_{\alpha\beta}\;
	{\cal E}^l_{pq}(\beta,\alpha;\uk)_\mathrm{E,M}\;,
\end{equation}
where $\bm{\sigma}_i$, $i=0,1,2,3$, are the Pauli matrices, with
$\bm{\sigma}_0$ the 2$\times$2 identity. Again, we follow the convention
adopted by \citetalias{LL04} for the definition of the Pauli matrices 
(see eq.~[5.128]).
The polarization tensors ${\cal T}^l_{pq}$ satisfy the orthogonality property
\begin{equation} \label{eq:Tortho}
\sum_{pq} {\cal T}^l_{pq}(i,\uk)_\mathrm{E,M}^\ast\,
{\cal T}^l_{pq}(j,\uk)_\mathrm{E,M}=\frac{1}{2}\,\delta^i_j\;,
\end{equation}
for all orders $l$.
This follows immediately from the definitions (\ref{eq:Edef}) and the 
trace property (\ref{eq:Etens_contr0}), for which
\begin{eqnarray*}
\sum_{pq}
	{\cal E}^l_{pq}(\alpha,\beta;\uk)_\mathrm{E,M}^\ast\,
	{\cal E}^l_{pq}(\gamma,\delta;\uk)_\mathrm{E,M}
=\sum_{pq}
	{\cal E}^l_{pp}(\alpha,\gamma;\uk)_\mathrm{E,M}^\ast\,
	{\cal E}^l_{qq}(\beta,\delta;\uk)_\mathrm{E,M}
=\delta^\gamma_\alpha\,\delta^\delta_\beta\;, 
\quad\forall l\in \mathrm{N}\;,
\end{eqnarray*}
and from the orthogonality of the Pauli matrices,
\begin{displaymath}
\sum_{\alpha\beta=\pm 1} 
	(\bm{\sigma}_i^\ast)_{\alpha\beta}
	(\bm{\sigma}_j)_{\alpha\beta}=2\,\delta^j_i\;.
\end{displaymath}

Finally, through the polarization tensors ${\cal T}^l_{pq}$ 
and the Stokes parameters $S_i$ of the radiation field,
we can introduce the radiation tensors of rank 
$2l+1$ (cf.~\citetalias{LL04}, 
eq.~[5.153]; also App.~\ref{sec:Stokes})
\begin{equation} \label{eq:reduc_J}
J^l_{pq}(\omega)_\mathrm{E,M}=\sum_{i=0}^3 \oint\frac{d\uk}{4\pi}\;
{\cal T}^l_{pq}(i,\uk)_\mathrm{E,M}\, S_i(\omega,\uk)\;.
\end{equation}
These reducible tensors generalize the usual ones of rank $3$ that 
enter the expressions of the transition and relaxation rates for 
radiative processes in the statistical equilibrium of electric 
(and magnetic) dipole transitions (see \citetalias{LL04}, Ch.~7).

\begin{deluxetable}{ll}
\tablewidth{0pt}
\tablecaption{\label{tab:geom_tens}
The irreducible spherical polarization tensors 
${\cal T}^{2;K}_Q(i,\uk)$, with $i=0,1,2,3$ standing for 
the four Stokes parameters $I,Q,U,V$. 
This table only lists the non-vanishing components 
with $Q\ge 0$ for 
E2 transitions.  
The components with $Q<0$ are derived according to
${\cal T}_{-Q}^{2;K}(i,\uk)=(-1)^Q\,{\cal T}_Q^{2;K}(i,\uk)^\ast$. 
The case of 
M2 transitions follows
from the relations (\ref{eq:Ttens_EvM}). 
Here $\uk\equiv(\vartheta,\varphi)$ is the 
propagation direction, while $\gamma$ specifies
the reference direction of linear polarization 
(see Fig.~\ref{fig:scatgeom}).}
\startdata
\\
${\cal T}^{2;0}_0(0,\uk)=1$
	& \\[12pt]
${\cal T}^{2;2}_0(0,\uk)=-\frac{1}{2}\sqrt{\frac{5}{14}}\,(3\cos^2\vartheta-1)$
	& ${\cal T}^{2;2}_0(1,\uk)=-\frac{3}{2}\sqrt{\frac{5}{14}}\,\cos 2\gamma
    	\sin^2\vartheta$ \\[3pt]
${\cal T}^{2;2}_1(0,\uk)=\frac{1}{2}\sqrt{\frac{15}{7}}\,\cos\vartheta\sin\vartheta\;
	\textrm{e}^{\imag\varphi}$
	& ${\cal T}^{2;2}_1(1,\uk)=-\frac{1}{2}\sqrt{\frac{15}{7}}\,
	(\cos 2\gamma\cos\vartheta+\imag
	\sin 2\gamma)\sin\vartheta\;{\rm e}^{\imag\varphi}$ \\[3pt]
${\cal T}^{2;2}_2(0,\uk)=-\frac{1}{4}\sqrt{\frac{15}{7}}\sin^2\vartheta\;
	\textrm{e}^{\imag 2\varphi}$
	& ${\cal T}^{2;2}_2(1,\uk)=-\frac{1}{4}\sqrt{\frac{15}{7}}\,
    	[\cos 2\gamma\,(1+\cos^2\vartheta)+\imag\,
    	2\sin 2\gamma \cos\vartheta]\,
    	{\rm e}^{\imag 2\varphi}$ \\[12pt]
${\cal T}^{2;4}_0(0,\uk)=\frac{1}{2\sqrt{14}}\,
	[5\sin^2\vartheta\,(7\cos^2\vartheta+1)-8]$
	& ${\cal T}^{2;4}_0(1,\uk)=\frac{5}{2\sqrt14}
	\cos2\gamma\,(7\cos^2\vartheta-1)
	\sin^2\vartheta$ \\[3pt]
${\cal T}^{2;4}_1(0,\uk)=\sqrt{\frac{5}{14}}\,
	(7\cos^2\vartheta-3)\cos\vartheta\sin\vartheta\,
	{\rm e}^{\imag \varphi}$
	& ${\cal T}^{2;4}_1(1,\uk)=-\frac{1}{2}\sqrt{\frac{5}{14}}\,
	[\cos2\gamma\,\cos\vartheta\,(7\sin^2\vartheta-3)$ \\[3pt]
& $\hphantom{{\cal T}^{2;4}_1(1,\uk)=-\frac{1}{2}\sqrt{\frac{5}{14}}\,}
	-\imag\sin2\gamma\,
	(7\cos^2\vartheta-1)]\sin\vartheta\,
	{\rm e}^{\imag\varphi}$ \\[3pt]
${\cal T}^{2;4}_2(0,\uk)=-\frac{1}{2}\sqrt{\frac{5}{7}}\,
	(7\cos^2\vartheta-1)\sin^2\vartheta\,
	{\rm e}^{\imag 2\varphi}$
	& ${\cal T}^{2;4}_2(1,\uk)=-\frac{1}{2}\sqrt{\frac{5}{7}}\,
	\bigl\{
	\cos2\gamma\,[\cos^2\vartheta\,(7\sin^2\vartheta-1)-1]$ \\[3pt]
& $\hphantom{{\cal T}^{2;4}_2(1,\uk)=-\frac{1}{2}\sqrt{\frac{5}{7}}\,}
	+\imag\sin2\gamma \cos\vartheta\,
	(7\sin^2\vartheta-2)\bigr\}\,
	{\rm e}^{\imag 2\varphi}$ \\[3pt]
${\cal T}^{2;4}_3(0,\uk)=\sqrt{\frac{5}{2}}\,
	\cos\vartheta\sin^3\vartheta\,
	{\rm e}^{\imag 3\varphi}$
	& ${\cal T}^{2;4}_3(1,\uk)=-\frac{1}{2}\sqrt{\frac{5}{2}}\,
	[2\cos2\gamma \cos^3\vartheta
	+\imag\sin2\gamma\,
	(3\cos^2\vartheta-1)]\sin\vartheta\,
	{\rm e}^{\imag 3\varphi}$ \\[3pt]
${\cal T}^{2;4}_4(0,\uk)=-\frac{\sqrt5}{4}\,
	\sin^4\vartheta\,
	{\rm e}^{\imag 4\varphi}$
	& ${\cal T}^{2;4}_4(1,\uk)=\frac{\sqrt5}{4}\,
	[\cos2\gamma\,(1+\cos^2\vartheta)
	+\imag\,2\sin2\gamma \cos\vartheta]\sin^2\vartheta\,
	{\rm e}^{\imag 4\varphi}$ \\[6pt]
\hline\\[-6pt]
${\cal T}^{2;1}_0(3,\uk)=\frac{1}{\sqrt2}\cos\vartheta$
	& \\[3pt]
${\cal T}^{2;1}_1(3,\uk)=-\frac{1}{2}\sin\vartheta\;
	{\rm e}^{\imag\varphi}$
	&  \\[12pt]
${\cal T}^{2;3}_0(3,\uk)=-\frac{1}{2\sqrt2}\,(5\cos2\vartheta-1)\cos\vartheta$
	& ${\cal T}^{2;K}_Q(2,\uk)={\cal T}^{2;K}_Q(1,\uk)\;
	\left\{\sin2\gamma\to\cos2\gamma\,,\;
	\cos2\gamma\to-\sin2\gamma\right\}$ \\[3pt]
${\cal T}^{2;3}_1(3,\uk)=\frac{1}{4}\sqrt{\frac{3}{2}}\,
	(5\cos2\vartheta+3)\sin\vartheta\,
	{\rm e}^{\imag \varphi}$
	&  \\[3pt]
${\cal T}^{2;3}_2(3,\uk)=-\frac{\sqrt15}{2}\cos\vartheta\sin^2\vartheta\,
	{\rm e}^{\imag 2\varphi}$
	& \\[3pt]
${\cal T}^{2;3}_3(3,\uk)=\frac{1}{2}\sqrt{\frac{5}{2}}\,
	\sin^3\vartheta\,
	{\rm e}^{\imag 3\varphi}$
	& \\[6pt]
\enddata
\end{deluxetable}
\vspace{-12pt}

The transformation of the reducible tensors ${\cal T}^l_{pq}$ 
into their irreducible counterparts ${\cal T}^{l:K}_Q$
is given by the usual relation (cf.~\citetalias{LL04},
eq.~[5.124]), apart from a normalization factor that can 
be arbitrarily set; see discussions in \citetalias{VMK88}, 
Sect.~3.1, or \citetalias{LL04}, after eq.~[2.79]),
\begin{equation} \label{eq:TlKQ}
{\cal T}^{l:K}_Q
=\sum_{pq} (-1)^{l+p}\,[(2l+1)(2K+1)]\shalf\,
\thrj{l}{l}{K}{p}{-q}{-Q}\,{\cal T}^l_{pq}\;,
\qquad K=0,\ldots,2l\;, 
\end{equation}
together with the inverse relation
\begin{equation} \label{eq:Tlpq}
{\cal T}^{l}_{pq}
=\sum_{KQ} (-1)^{l+p}\left(\frac{2K+1}{2l+1}\right)\shalf
\thrj{l}{l}{K}{p}{-q}{-Q}\,{\cal T}^{l:K}_Q\;,
\end{equation}
Our choice of normalization, slightly different from the usual one
because of the presence of the $(2l+1)\shalf$ factor, has the purpose 
of making 
${\cal T}^{l:0}_0(0,\uk)_\mathrm{E,M}=1$ for all $l$. With the definition (\ref{eq:TlKQ}), the
orthogonality relation (\ref{eq:Tortho}) can be extended to the
irreducible tensors as well. In fact, through simple Racah algebra, 
we find
\begin{equation} \label{eq:TorthoKQ}
\sum_{KQ} {\cal T}^{l:K}_Q(i,\uk)_\mathrm{E,M}^\ast\,
{\cal T}^{l:K}_Q(j,\uk)_\mathrm{E,M}=\frac{2l+1}{2}\,\delta^i_j\;.
\end{equation}
In strict analogy to eq.~(\ref{eq:TlKQ}), we introduce the tensors
\begin{equation} \label{eq:ElKQ}
{\cal E}^{l:K}_Q(\alpha,\beta;\uk)_\mathrm{E,M}
=\sum_{pq} (-1)^{l+p}\,[(2l+1)(2K+1)]\shalf\,
\thrj{l}{l}{K}{p}{-q}{-Q}\,{\cal E}^l_{pq}(\alpha,\beta;\uk)_\mathrm{E,M}\;,
\end{equation}
in terms of which, recalling eq.~(\ref{eq:Ttens}), we can write directly
\begin{equation} \label{eq:TlKQ.alt}
{\cal T}^{l:K}_{Q}(i,\uk)_\mathrm{E,M}=\frac{1}{2}\sum_{\alpha\beta=\pm1}
	(\bm{\sigma}_i)_{\alpha\beta}\;
	{\cal E}^{l:K}_{Q}(\beta,\alpha;\uk)_\mathrm{E,M}\;.
\end{equation}
The numerical implementation of eq.~(\ref{eq:ElKQ}) is made easier
if we rewrite it taking advantage of the definitions (\ref{eq:Edef}) 
and the summation properties of rotation matrices. In particular, for 
the case of 
E$l$ transitions, we get
\begin{eqnarray} \label{eq:ElKQ.num}
{\cal E}^{l:K}_Q(\alpha,\beta;\uk)_E
&=& (2l+1)\shalf
	\sum_{pq} (-1)^{l+q}\,(lp,l\,{-}q|KQ)\,
	D^l_{\alpha p}(\hat{R})\,D^l_{\beta q}(\hat{R})^\ast \nonumber \\
&=& (2l+1)\shalf
	\sum_{pq} (-1)^{l+q}\,(-1)^{\beta-q}\,(lp,l\,{-}q|KQ)\,
	D^l_{\alpha p}(\hat{R})\,D^l_{{-}\beta\,{-}q}(\hat{R}) \nonumber \\
&=& (-1)^{l+\beta}\,(2l+1)\shalf\,
	(l\alpha,l\,{-}\beta|K\,\alpha{-}\beta)\,
	D^K_{\alpha{-}\beta\,Q}(\hat{R}) \nonumber \\
&\equiv& (-1)^{l+\alpha}\,[(2l+1)(2K+1)]\shalf\,
\thrj{l}{l}{K}{\alpha}{-\beta}{\beta-\alpha}\,
	D^K_{\alpha{-}\beta\,Q}(\hat{R})\;.
\end{eqnarray}
The case of 
M$l$ transitions evidently yields exactly
the same expression, apart from an additional factor $\alpha\beta$.

Finally, using eq.~(\ref{eq:TlKQ.alt}), we can introduce the irreducible form of the 
radiation tensors, in strict analogy with eq.~(\ref{eq:reduc_J}),
\begin{equation} \label{eq:irreduc_J}
J^{l:K}_{Q}(\omega)_\mathrm{E,M}=\sum_{i=0}^3 \oint\frac{d\uk}{4\pi}\;
{\cal T}^{l:K}_{Q}(i,\uk)_\mathrm{E,M}\, S_i(\omega,\uk)\;.
\end{equation}
This expression will be used in the following section to calculate examples 
of the 
E2 radiation tensors in the solar atmosphere.

We want to conclude this presentation of the polarization tensors by
pointing out their relation between the electric- and magnetic-multipole 
cases. From eqs.~(\ref{eq:Ttens}) and (\ref{eq:TlKQ}), and the definitions
(\ref{eq:Edef}), it follows immediately that
\begin{eqnarray*}
{\cal T}^l_{pq}(i,\uk)_\mathrm{M}
&=&\frac{1}{2}\sum_{\alpha\beta=\pm1}
	\alpha\beta(\bm{\sigma}_i)_{\alpha\beta}\;
	{\cal E}^l_{pq}(\beta,\alpha;\uk)_\mathrm{E}\;, \\
{\cal T}^{l:K}_{Q}(i,\uk)_\mathrm{M}
&=&\frac{1}{2}\sum_{\alpha\beta=\pm1}
	\alpha\beta(\bm{\sigma}_i)_{\alpha\beta}\;
	{\cal E}^{l:K}_{Q}(\beta,\alpha;\uk)_\mathrm{E}\;.
\end{eqnarray*} 
The matrices
$\alpha\beta(\bm{\sigma}_i)_{\alpha\beta}$
are identical to the Pauli matrices for $i=0,3$, since they are purely
diagonal (hence $\alpha\beta=1$), but change sign for $i=1,2$, since the Pauli
matrices are purely off-diagonal in those cases (hence
$\alpha\beta=-1$). Thus, the following general relations are true for
every multipole order $l$ (cf.~\citetalias{LL04}, \S6.8),
\begin{subequations}
\label{eq:Ttens_EvM}
\begin{alignat}{3}
&{\cal T}^l_{pq}(i,\uk)_\mathrm{M}={\cal T}^l_{pq}(i,\uk)_\mathrm{E}\;,\qquad
&&{\cal T}^{l:K}_{Q}(i,\uk)_\mathrm{M}={\cal
T}^{l:K}_{Q}(i,\uk)_\mathrm{E}\;,\qquad
&&i=0,3 \\
&{\cal T}^l_{pq}(i,\uk)_\mathrm{M}=-{\cal T}^l_{pq}(i,\uk)_\mathrm{E}\;,\qquad
&&{\cal T}^{l:K}_{Q}(i,\uk)_\mathrm{M}=-{\cal
T}^{l:K}_{Q}(i,\uk)_\mathrm{E}\;.\qquad
&&i=1,2
\end{alignat}
\end{subequations}
Table~\ref{tab:geom_tens} gives the expressions of ${\cal T}^{2:K}_Q(i,\uk)$ for 
the 
E2 case. The expressions for the 
E1 case 
are tabulated by \citetalias{LL04} (Table~5.6).

At a glance, the comparison
between the dipolar and quadrupolar tensor components in
Table~\ref{tab:geom_tens} shows that, apart from a numerical 
factor, the shared multipoles $K=0,1,2$ take exactly the same 
geometric form. In contrast, the quadrupolar tensors carry a more 
convoluted geometric dependence for the new multipoles 
$K=3,4$, which will manifest itself through new signatures of 
scattering polarization (see, e.g., Fig.~\ref{fig:hanleplots} below). 

\subsection{Application: spontaneous emission rates}
\label{sec:R_E}

As a straightforward application of the derived formalism, we evaluate 
the relaxation rate for spontaneous emission (s.e.) 
from an excited state $m$
towards all lower states $n\prec m$ radiatively bound to $m$. This rate
is derived from first principles (e.g., \citetalias{LL04}, from
eq.~[6.48] onward; also \citealt{La83}), 
and using the notation of this paper is given by
\begin{equation} \label{eq:R_E_def}
R(m)_{\rm s.e.}=\frac{1}{\hbar^2}\sum_{n\prec m}\sum_{\vk}
|Q(\omega,\uk,\beta)_{mn}|^2\,\bigl[\Phi_{nm}(\omega)+\Phi_{nm}^\ast(\omega)\bigr]\;,
\end{equation}
where
$\Phi_{nm}(\omega)=[\epsilon_m+\epsilon_n-\imag(\omega_{mn}-\omega)]^{-1}$ is
the spectral profile of the transition,\footnote{The line
profile arises from the time integral in the master equation
of the interacting system (e.g., \citetalias{LL04}, Ch.~6, p.~255).}
 corresponding to the natural line 
width due to the inverse lifetimes $\epsilon_m$ and
$\epsilon_n$ of the transition levels
\cite[cf.][{eq.~[34]}]{Ca14}. Here, $Q(\omega,\uk,\beta)_{mn}$ 
is one of the matrix elements (\ref{eq:qmat1.summary}) or
(\ref{eq:mmat1.summary}).
Further, the sum over all possible radiation modes $\vk$ implies the following
transition to the continuum,\footnote{We note that this yields a 
logarithmically divergent integral of the line profile, which can be
treated using standard renormalization techniques, ultimately leading to
$(1/\pi)\int_0^\infty d\omega\;\omega
\left[\Phi_{nm}(\omega)+\Phi_{nm}^\ast(\omega)\right]\approx
2\,\omega_{mn}\approx 2\,\omega$.}
\begin{equation} \label{eq:to_cont}
\sum_{\vk} \longrightarrow
        \frac{1}{2\pi^2}\frac{\cal V}{c^3} 
        \int_0^\infty d\omega\;\omega^2 \oint \frac{d\uk}{4\pi}\,
        \sum_\beta\;.
\end{equation}
For an electric- or magnetic-multipole transition of order $l$, after
substitution of either one of eqs.~(\ref{eq:qmat1.summary}) or
(\ref{eq:mmat1.summary})
into eq.~(\ref{eq:R_E_def}), and using the property 
(\ref{eq:Etens_contraction}), we find
\begin{equation} \label{eq:R_E^l}
R^l(m)_{\rm s.e.}=\frac{2}{[(2l-1)!!]^2}\,
	\frac{l+1}{2l}\,\frac{2}{2l+1}\,
	\frac{\omega^{2l+1}}{\hbar c^{2l+1}}
	\sum_{n\prec m}\sum_q |\bra{m}\mathscr{O}^l_q\ket{n}|^2\;,
\end{equation}
where $\mathscr{O}^l_q=\{\mathscr{Q}^l_q,\mathscr{M}^l_q\}$. 
Specifically for dipolar, quadrupolar, and 
octupolar interactions, we find, respectively,
\begin{eqnarray*}
R^1(m)_{\rm s.e.}
&=&\frac{4}{3}\,\frac{\omega^3}{\hbar c^3}
	\sum_{n\prec m}\sum_q |\bra{m}\mathscr{O}^1_q\ket{n}|^2\;, \\
R^2(m)_{\rm s.e.}
&=&\frac{1}{15}\,\frac{\omega^5}{\hbar c^5}
	\sum_{n\prec m}\sum_q |\bra{m}\mathscr{O}^2_q\ket{n}|^2\;, \\
R^3(m)_{\rm s.e.}
&=&\frac{8}{4725}\,\frac{\omega^7}{\hbar c^7}
	\sum_{n\prec m}\sum_q |\bra{m}\mathscr{O}^3_q\ket{n}|^2\;,
\end{eqnarray*}
in agreement with the results of \citetalias{SM68}, \S10.6-7.

\section{Hanle effect of electric-quadrupole transitions in a
two-level atom} \label{sec:hanle}

To demonstrate the formalism developed in Sect.~\ref{sec:genform}, 
we model the Hanle effect of the scattering polarization from a E2 
transition in the presence of a magnetic field. 

We consider a collisionless plasma of two-level atoms, 
illuminated by an unpolarized radiation field with cylindrical 
symmetry around the local vertical through the scattering point
(Fig.~\ref{fig:scatgeom}). 
We assume that we can neglect the effects of stimulated emission,
as well as the polarization of the lower level $J_l$. Under these
conditions, we can adapt 
the expression (10.27) of \citetalias{LL04} for the density matrix 
of the upper level $J_u$, expressed in the magnetic reference frame, 
$B$, having the $z$-axis (quantization axis) along the $\bm{B}$ vector
($z'$-axis in Fig.~\ref{fig:scatgeom}).
Thus, we can write for the normalized density matrix tensor
$\varrho^K_Q(J_u)=\rho^K_Q(J_u)/\rho^0_0(J_l)$,
\begin{eqnarray} \label{eq:rhoKQ}
\varrho^K_Q(J_u)
&=&\left(\frac{2J_l+1}{2J_u+1}\right)\shalf
	\frac{B_{J_l J_u}}{A_{J_u J_l}+\imag\,Q\,g_{J_u} \omega_B}\,
	w\apx{2:$K$}_{J_u J_l}\,
	(-1)^Q\,J^{2:K}_{-Q}(\omega_{ul})_B\;,
\qquad K=0,\ldots,2J_u\;,
\end{eqnarray}
where 
$A_{J_u J_l}$ and $B_{J_l J_u}$ are, respectively, 
the Einstein coefficients for spontaneous emission and 
absorption of the E2 transition,
$\omega_B=eB/(2mc)$ is the Larmor frequency corresponding to 
the field strength $B$, $g_{J_u}$ is the Land\'e factor of the 
excited level, and finally we introduced the polarizability factor 
of the E2 transition
\begin{equation} \label{eq:polarizability}
w\apx{2:$K$}_{J_u J_l}=\sixj{2}{2}{K}{J_u}{J_u}{J_l}\biggl/
		    \sixj{2}{2}{0}{J_u}{J_u}{J_l}\;,\qquad K=0,...,4\;.
\end{equation}
We note that the non-nullity condition of the polarizability factor,
automatically restricts the highest possible multipole order $K$ in
eq.~(\ref{eq:rhoKQ}) to $\min\{4,2J_u\}$.

Because the incident radiation field is assumed to be unpolarized and 
cylindrically symmetric around the local vertical, the radiation 
tensors in the $B$-frame entering eq.~(\ref{eq:rhoKQ}) are simply 
given by 
\begin{equation} \label{eq:RadJ_B}
J^{2:K}_Q(\omega_{ul})_B=D^K_{0Q}(\hat{R}_{BS})\,J^{2:K}_0(\omega_{ul})_S\;,
\end{equation}
where $\hat{R}_{BS}$ is the rotation operator that transforms the
solar frame, $S$ ($z$-axis along the local vertical; see
Fig.~\ref{fig:scatgeom}) into the $B$-frame. In addition, because 
of the assumed symmetry, all \emph{odd} 
multipole orders of the radiation field are identically zero
(cf.~eq.~[\ref{eq:irreduc_J}] and Table~\ref{tab:geom_tens}). This
property extends directly to the density matrix solution, as seen from
eq.~(\ref{eq:rhoKQ}).

The density matrix solution (\ref{eq:rhoKQ}) can be used to 
derive the spectrally integrated, fractional linear\footnote{The 
spectrally integrated circular polarization is
always vanishing under the conditions of this example.} 
polarization of the emission line, 
$p_i(\uk)\equiv\varepsilon_i(\uk)/\varepsilon_0(\uk)$, for $i=1,2$.
Using eq.~(10.31) of \citetalias{LL04}, and eqs.~(\ref{eq:rhoKQ}) and
(\ref{eq:RadJ_B}) above, we can write, for $i=0,1,2$,
\begin{eqnarray} \label{eq:epsKQ}
\varepsilon_i(\uk)
&\propto&\sum_{KQ}
	w\apx{2:$K$}_{J_u J_l}\;{\cal T}^{2:K}_Q(i,\uk)_B\,
	\frac{\varrho^K_Q(J_u)}{\varrho^0_0(J_u)} \nonumber \\
&\propto&\sum_{KQ}
	\Bigl[w\apx{2:$K$}_{J_u J_l}\Bigr]^2\,
	{\cal T}^{2:K}_Q(i,\uk)_B\,
	\frac{(-1)^Q\,D^K_{0\,-Q}(\hat R_{BS})}%
	{1+\imag\,Q\,g_{J_u}\omega_B/A_{J_uJ_l}}\,
	\frac{J^{2:K}_0(\omega_{ul})_S}{J^{2:0}_0(\omega_{ul})_S}
	\nonumber \\
&\propto&\sum_{K}
	\Bigl[w\apx{2:$K$}_{J_u J_l}\Bigr]^2
	\frac{J^{2:K}_0(\omega_{ul})_S}{J^{2:0}_0(\omega_{ul})_S}
	\sum_P {\cal T}^{2:K}_P(i,\uk)_S \sum_Q
	\frac{(-1)^Q\,D^K_{PQ}(\hat{R}_{BS})\,D^K_{0\,-Q}(\hat R_{BS})}%
	{1+\imag\,Q\,g_{J_u}\omega_B/A_{J_uJ_l}}\;,
\end{eqnarray}
where in the last line we used
\begin{equation} \label{eq:Tpol_B}
{\cal T}^{2:K}_Q(i,\uk)_B=
	\sum_P D^K_{PQ}(\hat{R}_{BS})\,
	{\cal T}^{2:K}_P(i,\uk)_S\;.
\end{equation}
%
%
If we indicate with $\bm{\hat b}\equiv(\vartheta_B,\varphi_B)$ the direction of the
magnetic field in the solar frame, we may choose the magnetic frame 
$B\equiv(O;x'y'z')$ such that the transformation $\hat{R}_{BS}$ consists only 
of a rotation by $\varphi_B$ around the local vertical, followed by a rotation 
by $\vartheta_B$ around the transformed $y$-axis (see
Fig.~\ref{fig:scatgeom}). We then have,
\begin{displaymath}
(-1)^Q\,D^K_{0\,-Q}(\hat{R}_{BS})=d^K_{0Q}(\vartheta_B)\;,\qquad
D^K_{PQ}(\hat{R}_{BS})=\exp(-\imag\,P\varphi_B)\,d^K_{PQ}(\vartheta_B)\;,
\end{displaymath}
where $d^K_{PQ}(\beta)$ is the reduced rotation matrix
(\citetalias{VMK88}, \S4.3). Thus, eq.~(\ref{eq:epsKQ})
becomes explicitly,
\begin{equation} \label{eq:epsKQ.expl}
\varepsilon_i(\uk)
\propto\sum_K \Bigl[w\apx{2:$K$}_{J_u J_l}\Bigr]^2\,
\frac{J^{2:K}_0(\omega_{ul})_S}{J^{2:0}_0(\omega_{ul})_S}
\sum_P\,{\cal T}^{2:K}_P(i,\uk)_S\,\exp(-\imag\,P\varphi_B)
\sum_Q \frac{d^K_{PQ}(\vartheta_B)\,d^K_{0Q}(\vartheta_B)}%
{1+\imag\,Q\,g_{J_u}\omega_B/A_{J_uJ_l}}\;.\qquad (i=0,1,2)
\end{equation}

A typical system described by such a model (apart from the
additional assumption of negligible stimulated emission) is 
the transition $(J_l,J_u)\equiv(0,2)$, for which
\begin{equation}
w\apx{2:2}_{20}=w\apx{2:4}_{20}=1\;.
\end{equation}
For a transition $(J_l,J_u)\equiv(1/2,5/2)$, which also has a
non-polarizable lower level by radiation anisotropy, we have instead
\begin{equation}
w\apx{2:2}_{\frac{5}{2}\,\frac{1}{2}}=2/\sqrt5\;,\qquad
w\apx{2:4}_{\frac{5}{2}\,\frac{1}{2}}=1/\sqrt3\;.
\end{equation}
Both types of transitions require the calculation of the radiation 
anisotropies in the solar frame for the orders $K=2,4$, which we 
undertake in the next section (see eqs.~[\ref{eq:W1-W2}] below).

\subsection{Radiation anisotropy tensors}
\label{sec:radan}

Here we explicitly compute the radiation tensors $J^{l:K}_Q(\omega)_S$ 
generalized to the multipole order $K=4$, which are necessary to describe 
the scattering polarization in the 
E2 transition 
$(J_l,J_u)\equiv(0,2)$.
Because we are considering the case of unpolarized incident radiation,
we remark that the following derivation applies identically to both 
E2 and M2 cases. 

%

While the incident radiation is further assumed to be cylindrically symmetric 
around the local vertical to the solar surface, we are taking into
account the presence of a wavelength-dependent center-to-limb variation (CLV), 
represented through the usual polynomial expression of \cite{Al73},
\begin{equation} \label{eq:clv}
I(\lambda,\mu)=I_0(\lambda)\left[1-u(\lambda)(1-\mu)-v(\lambda)(1-\mu^2)\right]\;,
\qquad \mu=\cos\vartheta\;,
\end{equation}
where $\vartheta$ is the angle between
the line-of-sight and the local vertical through the observed point (see
Fig.~\ref{fig:scatgeom}).
Following \citeauthor{La82} (\citeyear{La82}; cf.\ also 
\citetalias{LL04}, \S12.3), the irreducible radiation tensors 
relevant to dipolar transitions can be written in the form
\begin{subequations}
\begin{eqnarray}
J^{1:0}_0
&=&\frac{1}{2}\,I_0\left(a_1+a_2 u+a_3 v\right)\;, \\
J^{1:2}_0
&=&\frac{1}{4\sqrt2}\,I_0\left(b_1+b_2 u+b_3 v\right)\;,
\end{eqnarray}
\end{subequations}
and are obtained by integrating eq.~(\ref{eq:clv}) within the radiation cone 
between 0 (disk center) and $\vartheta_0$ (limb edge for a given height 
$h$, defined to be 1 at the solar surface; see Fig.~\ref{fig:scatgeom}), 
weighted by the proper polarization tensors ${\cal T}^{1:K}_0$ for the 
intensity (see eq.~[\ref{eq:TlKQ}]). We find (cf.~\citealt{La82})
\begin{subequations}
\begin{eqnarray}
a_1&=&1-\mu_0\;, \\
a_2&=&-a_1+\frac{1}{2}\biggl(1+\frac{\mu_0^2\,l_0}{\sqrt{1-\mu_0^2}}\biggr)\;, \\
a_3&=&\frac{1}{3}\biggl(\mu_0-\frac{2}{1+\mu_0}\biggr)\;,
\end{eqnarray}
\end{subequations}
and
\begin{subequations}
\begin{eqnarray}
b_1&=&\mu_0(1-\mu_0^2)\;, \\
b_2&=&-b_1-\frac{1}{8}\left[(3\mu_0^2-4)\biggl(1
	-\frac{\mu_0^2\,l_0}{\sqrt{1-\mu_0^2}}\biggr)+2\right]\;, \\
b_3&=&\frac{1}{15}\left[\mu_0(9\mu_0^2-11)+\frac{4}{1+\mu_0}\right]\;,
\end{eqnarray}
\end{subequations}
where we defined
\begin{equation}
\mu_0=\sqrt{1-\frac{1}{h^2}}\;,\qquad
l_0=\ln\frac{\mu_0}{1+\sqrt{1-\mu_0^2}}=\frac{1}{2}\ln\frac{h-1}{h+1}\;.
\end{equation}

\begin{figure}
\centering
\includegraphics[width=.85\textwidth,clip]{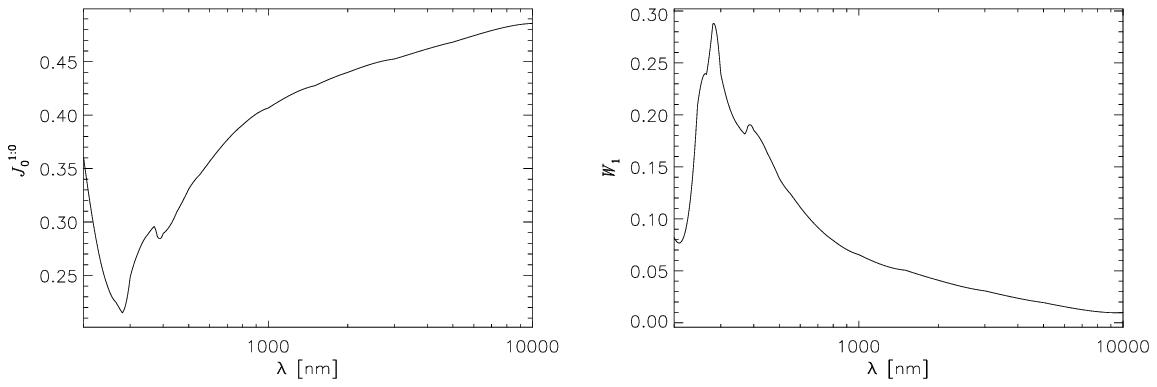}
\caption{\label{fig:E1_surf}
Plots of the $\displaystyle J^{1:0}_0(\lambda)$ radiation tensor (left) and radiation
anisotropy $W_1=\sqrt2\,J^{1:2}_0(\lambda)/J^{1:0}_0(\lambda)$ (right), 
between 200\,nm and 10\,$\mu$m, for $h=1\,R_\odot$ (solar surface),
adopting the CLV data of \citet{Al73}. These are the radiation
tensors necessary for the modeling of polarized line formation in
dipolar transition systems.}
\end{figure}

\begin{figure}
\centering
\includegraphics[width=.85\textwidth,clip]{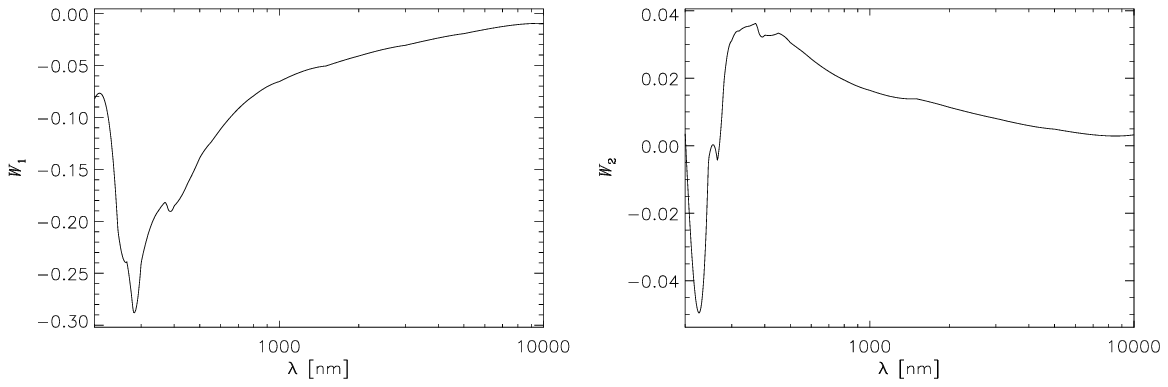}
\caption{\label{fig:E2_surf}
Plots of the radiation anisotropies $W_1$ (left) and $W_2$ (right), 
as defined by eqs.~(\ref{eq:W1-W2}). Atmospheric conditions and wavelength 
range are the same as for
Fig.~\ref{fig:E1_surf}. These radiation tensors need to be included, alongside 
with the dipolar tensors (noting the condition $J^{2:0}_0\equiv J^{1:0}_0$) in order 
to model the scattering polarization from 
E2 transitions in
multi-level atomic systems where the atomic levels are also populated via 
E1 transitions.} 
\end{figure}

\begin{figure}
\centering
\includegraphics[width=.495\textwidth]{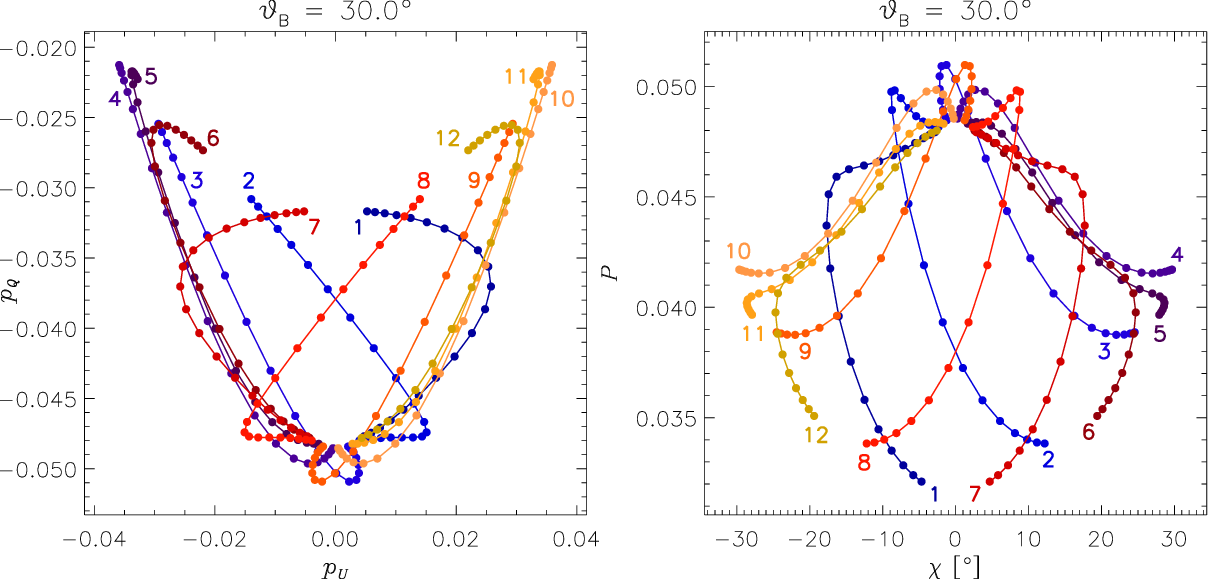}
\includegraphics[width=.495\textwidth]{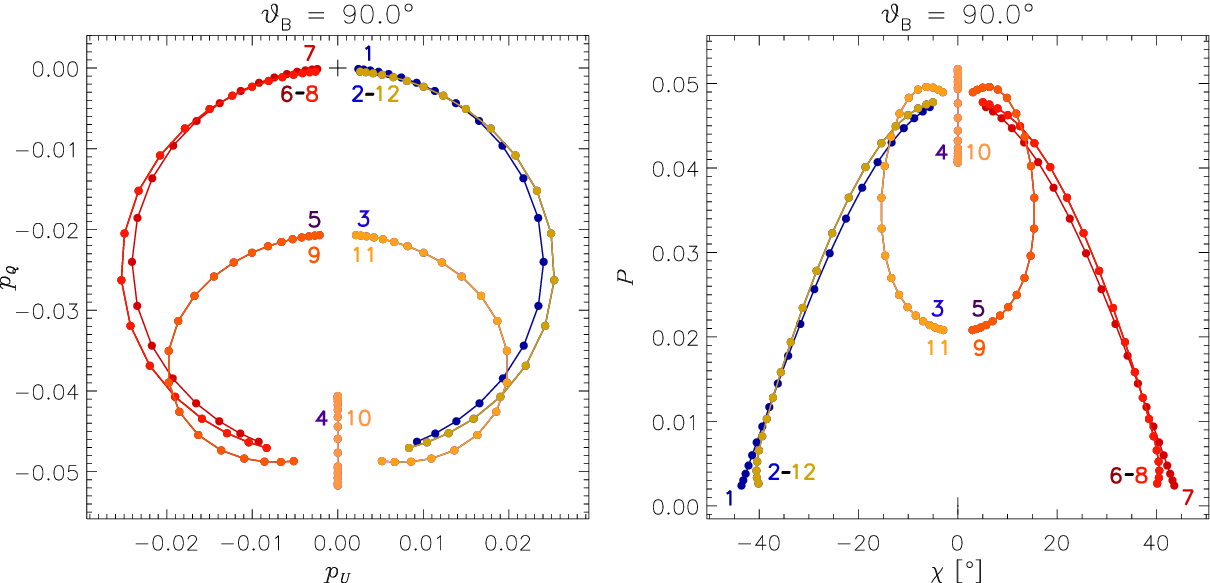}\vspace{3pt}
\includegraphics[width=.495\textwidth]{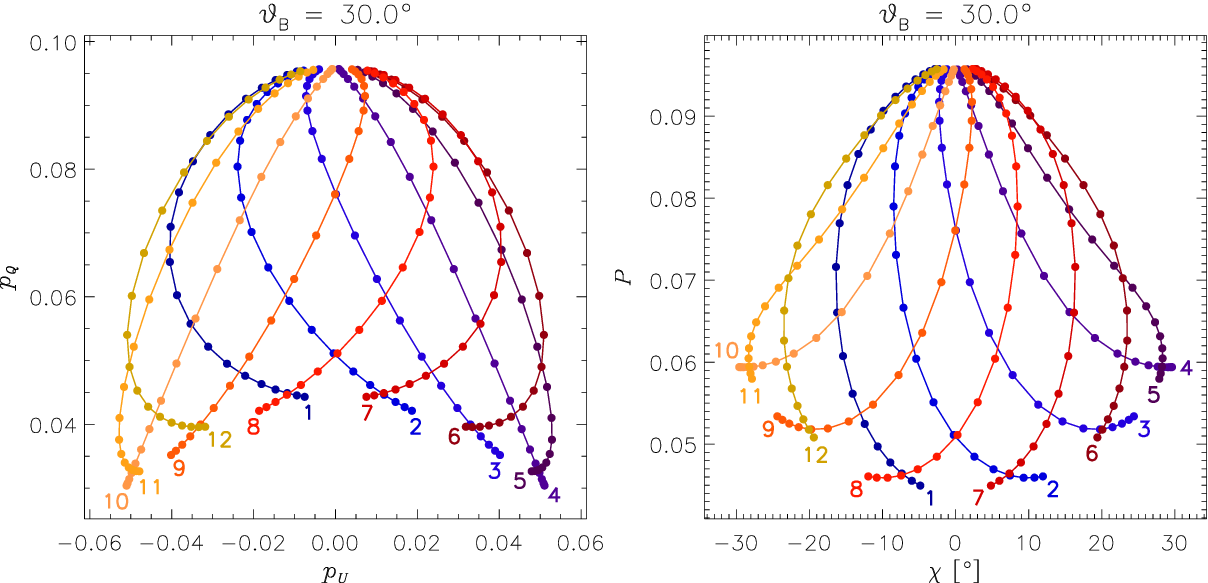}
\includegraphics[width=.495\textwidth]{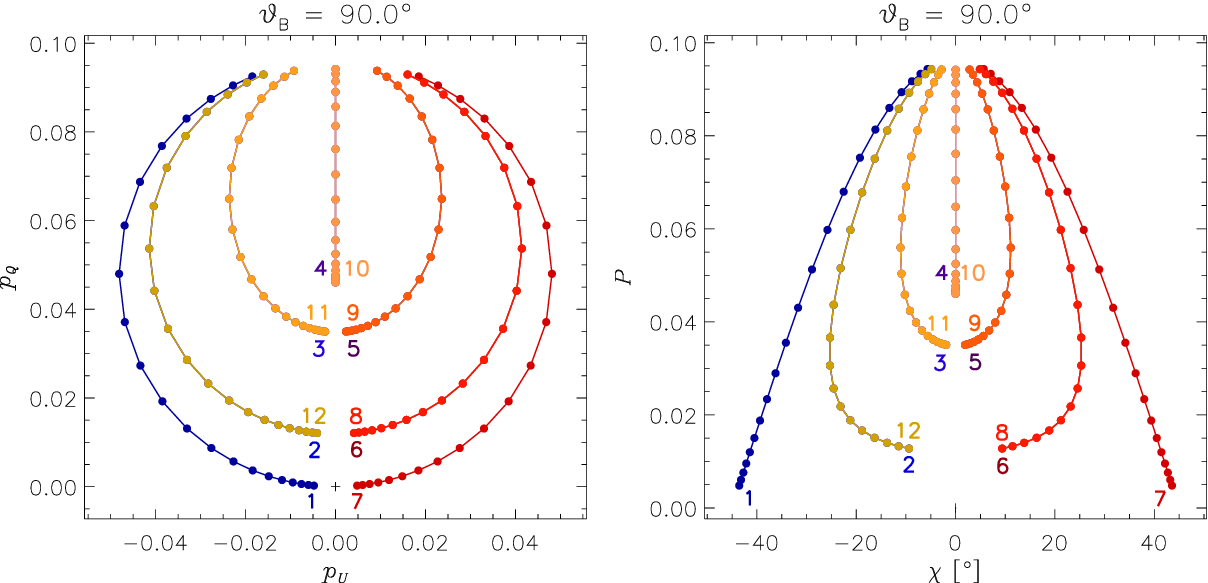}
\caption{\label{fig:hanleplots} Hanle diagrams for the 90$^\circ$
scattering of resonant radiation in the E2 transition 
$(J_l,J_u)\equiv(0,2)$ (top row) and in the 
E1 transition $(J_l,J_u)\equiv(0,1)$ (bottom row). 
We considered two inclinations of the magnetic field from 
the local vertical, $\vartheta_B=30^\circ$ (left column) 
and $\vartheta_B=90^\circ$ (right column). 
For each of the four plots, the Hanle diagrams are given 
as parametric curves in both forms of $(p_Q,p_U)$ (left panels) 
and $(P,\chi)$ (right panels), where $P=(p_Q^2+p_U^2)^{1/2}$ and 
$\chi=(1/2)\arctan(p_U/p_Q)$ (see text). 
The 12 colored curves are for different field azimuths 
$\varphi_B\in[-180^\circ,180^\circ)$ in steps of
$30^\circ$, and on each curve the symbols mark the field strength 
$\omega_B\in[0.1,10]\,A_{J_u J_l}$ in a decimal-log scale (21 points). 
The anisotropy conditions of the
incident radiation are given in the text.}
\end{figure}

We can proceed similarly to derive the radiation tensors relevant to
E2 transitions.
Because of the relations between the corresponding polarization tensors, 
\begin{displaymath} \textstyle
{\cal T}^{2:0}_0(i,\uk)={\cal T}^{1:0}_0(i,\uk)=\delta_{i0}\;,\qquad
{\cal T}^{2:2}_0(i,\uk)=-\sqrt{\frac{5}{7}}\;{\cal T}^{1:2}_0(i,\uk)\;,
\qquad i=0,1,2,3\;,
\end{displaymath}
we choose to define
\begin{subequations}
\begin{eqnarray}
J^{2:0}_0&=&\frac{1}{2}\,I_0\left(A_1+A_2 u+A_3 v\right)\;, \\
J^{2:2}_0&=&-\frac{1}{4}\sqrt{\frac{5}{14}}\,I_0\left(B_1+B_2 u+B_3 v\right)\;, \\
J^{2:4}_0&=&-\frac{1}{4\sqrt{14}}\,I_0\left(C_1+C_2 u+C_3 v\right)\;,
\end{eqnarray}
\end{subequations}
so we have at once
\begin{displaymath}
(A_1,A_2,A_3)\equiv(a_1,a_2,a_3)\;,\qquad
(B_1,B_2,B_3)\equiv(b_1,b_2,b_3)\;.
\end{displaymath}
Thus, we must only calculate $(C_1,C_2,C_3)$ anew, for which we find
\begin{subequations}
\begin{eqnarray}
C_1&=&\mu_0(1-\mu_0^2)(7\mu_0^2-3)\;, \\
C_2&=&\biggl( 7\mu_0^5-10\mu_0^3
	-\frac{35}{24}\,\mu_0^2+3\mu_0+\frac{4}{3} \biggr) 
	-\frac{1}{16}\,(35\mu_0^4-60\mu_0^2+24)\biggl(1
	-\frac{\mu_0^2\,l_0}{\sqrt{1-\mu_0^2}}\biggr)\;,\\
C_3&=&-\mu_0(1-\mu_0^2)(5\mu_0^2-3)\;.
\end{eqnarray}
\end{subequations}
We choose to define the 1st- and 2nd-order anisotropies for the
quadrupolar case as follows,
\begin{equation}
\label{eq:W1-W2}
W_1=\sqrt{\frac{14}{5}}\,\frac{J^{2:2}_0}{J^{2:0}_0}\;, \qquad
W_2=\frac{\sqrt{14}}{4}\,\frac{J^{2:4}_0}{J^{2:0}_0}\;,
\end{equation}
both of which have a lower asymptotic bound at $-1$ for $\mu_0\to 1$. In
particular, with these definitions, $W_1$ happens to be identical in magnitude 
but opposite in sign to the usual radiation anisotropy for the dipolar case.
This is demonstrated by the $W_1$ plots shown in
Figs.~\ref{fig:E1_surf} and \ref{fig:E2_surf}.
Consequently, we can anticipate that the linear polarization arising 
from an 
E2 transition will have the opposite sign with 
respect to an 
E1 transition formed under the same physical 
conditions, similarly
to what happens in the case of 
M1 transitions
(see, e.g., \citealt{CJ99}, \citetalias{LL04}). However, while in 
M1 transitions the polarization signatures only undergo a sign
change with respect to their 
E1 counterparts, in the case of 
E2 transitions, new polarization patterns will arise
as a consequence of the line polarizability $W_2$ shown in 
Fig.~\ref{fig:E2_surf}. This is clearly demonstrated by the examples 
discussed in the next section (cf.~Fig.~\ref{fig:hanleplots}).

\subsection{Scattering polarization and Hanle effect of the 2-0 atom}
\label{sec:scatpol}

As an application of these results, we model the scattering 
of polarized radiation in the 
E2 transition 
$(J_l,J_u)\equiv(0,2)$. 
We assume for simplicity a height $h=1$ ($\mu_0=0$) over the solar
surface, so the only radiation anisotropy comes from the CLV of the solar 
atmosphere. 
Because we assume an unpolarized radiation with cylindrical 
symmetry around the local vertical, all odd multipole orders of the 
density matrix solution vanish identically, so the corresponding 
summation in eq.~(\ref{eq:epsKQ.expl}) is limited to $K=0,2,4$. 
Additionally, we assume the scattering direction in the local-vertical 
frame of reference ($z$-axis along the local vertical, and $xz$-plane 
coinciding with the Sun's central meridian) to be 
$\uk\equiv(\vartheta,\varphi)\equiv(90^\circ,0)$. Finally, we assume the
tangent to the limb as the reference direction for positive linear 
polarization ($\gamma=90^\circ$). The tensors
${\cal T}^{2:K}_P(i,\uk)_S$ in eq.~(\ref{eq:epsKQ.expl}) can then be calculated
from Table~\ref{tab:geom_tens}.
%

With the above assumptions, we also have
\begin{displaymath} \textstyle
(A_1,A_2,A_3)\equiv(1,-\frac{1}{2},-\frac{2}{3})\;,\qquad
(B_1,B_2,B_3)\equiv(0,\frac{1}{4},\frac{4}{15})\;,\qquad
(C_1,C_2,C_3)\equiv(0,-\frac{1}{6},0)\;,\qquad
\end{displaymath}
and assuming a wavelength of 550\,nm, for which 
$u\simeq 0.93$ and $v\simeq -0.23$ according to \cite{Al73}, we derive
\begin{displaymath}
\frac{J^{2:2}_0}{J^{2:0}_0}\simeq-0.0743\;,\qquad
\frac{J^{2:4}_0}{J^{2:0}_0}\simeq 0.0301\;,
\end{displaymath}
or alternatively (cf.~Fig.~\ref{fig:E2_surf}),
\begin{displaymath}
W_1\simeq-0.1243\;,\qquad W_2\simeq 0.0281\;.
\end{displaymath}

Figure~\ref{fig:hanleplots} shows the Hanle polarization diagrams
calculated using eq.~(\ref{eq:epsKQ.expl}), with $p_Q=p_1(\uk)$ and
$p_U=p_2(\uk)$. 
We display both forms of these diagrams as parametric curves $(p_Q,p_U)$ 
(left panels) and $(P,\chi)$ (right panels), where 
$P=(p_Q^2+p_U^2)^{1/2}$ is the linear-polarization degree and 
$\chi=(1/2)\arctan(p_U/p_Q)$ gives the ``azimuth'' in the
plane-of-sky of the linear-polarization direction. 
Two different inclinations of the magnetic field from the local
vertical, $\vartheta_B=30^\circ,90^\circ$, are shown. The 12
colored curves in each plot
correspond to different field azimuths $\varphi_B$ in the interval 
$[-180^\circ,180^\circ)$ in steps of $30^\circ$. The points on each
curve correspond to different field strengths, such that
$\omega_B/A_{J_u J_l}$ varies in the interval $[0.1,10]$
in a decimal-log scale. The series of diagrams at the top are for the 
E2 transition $(J_l,J_u)\equiv(0,2)$ discussed in this application. For
comparison, the series of diagrams at the bottom show the corresponding
diagrams for the E1 transition $(J_l,J_u)\equiv(0,1)$, under exactly the
same physical conditions. 

Apart from the obvious difference in the sign of the scattering
polarization across the Hanle-sensitivity field regime---a
distinguishing characteristic also shared with M1 transitions
(cf.~eqs.~[\ref{eq:Ttens_EvM}])---the diagrams of Fig.~\ref{fig:hanleplots} 
demonstrate a
higher degree of degeneracy of the E2 polarization signature to the
magnetic field geometry, under similar observing conditions and
signal-to-noise ratios, especially for the frequently occurring case of
near right-angle scattering (panels to the right). A full exploration of
the diagnostic potential and limitations of these transitions is
therefore necessary, especially on how they can best be utilized
alongside with permitted transitions, which goes beyond the scope of
the formalism description of this work.

\section{Conclusions}

The extension to quadrupolar (E2) radiative transitions of 
the CRD theory of polarized line formation including non-isotropic 
excitation mechanisms
(as presented in \citetalias{LL04}) is an important addition to the
theoretical tools necessary to model resonance scattering polarization
in complex atoms. Such an extension becomes straightforward when we 
adopt the generalized tensorial formalism presented here. 
The first key ingredient for such a generalization is the 
adoption of the irreducible spherical representation of the  
tensorial quantities that describe the excitation state of the atomic
system (density matrix) and the interacting radiation field
(polarized radiation tensors), as presented by \citetalias{LL04}.
The second one, enabling this generalization to be extended to multipole
terms of arbitrary order, is the use of the spherical-wave expansion of 
the interaction Hamiltonian. 
This formalism can be utilized, for example, to extend the theoretical 
framework of the polarized line formation in complex atoms with both fine 
and hyperfine structures, presented by \cite{CM05}, to include E2 
transitions as well. Such an extension will be the subject of a future 
paper.

While the scattering polarization of magnetic-dipole (M1) 
transitions 
is a well-established diagnostics of the solar corona (see cited references
in the Introduction), 
E2 transitions 
in the solar spectrum are elusive, and their study may be presumed to be irrelevant 
for diagnostic purposes. The situation is dramatically different in colder and rarefied 
gas clouds such as the circumstellar environment of T-Tauri stars, or in
\ion{H}{2} regions. In these astrophysical objects,
the excitation state of relatively abundant atomic species, such as O, N, and S,
strongly departs from LTE, being completely dominated by the ambient
(and naturally anisotropic) radiation field from nearby stellar sources.
Because collisional processes are essentially negligible, forbidden 
radiative transitions---mostly 
of the M1 and E2 types---stand out in the spectra emitted by 
these objects. Thus, any polarimetric study aimed at diagnosing their magnetic 
properties, as well as to investigate possibly unresolved geometric structures 
(e.g., anisotropic stellar winds),
necessitates a proper modeling of the multipolar interactions, which
ultimately must be based on a self-consistent formalism such as the one
presented in this work. However, even in solar plasmas, forbidden
transitions can play a role in the statistical equilibrium of atomic
species that dominate the polarized solar spectrum, e.g., by relaxing 
the metastability of low-energy levels \cite[e.g.,][]{MSTB03}.

The simple examples of scattering polarization signals from 
E2 transitions presented in this work reveal striking 
differences with the much more familiar signatures from permitted 
transitions. These can be
regarded as new diagnostic opportunities but they also come with potential 
limitations that need to be thoroughly investigated. We hope that this
introductory work can open a new diagnostic window relying on the use of
the Hanle effect for the investigation of weakly magnetized plasmas in
all branches of astrophysics 
where forbidden transitions are routinely observed.

Multipolar transitions are also relevant to the laboratory
investigation of the quantum theory of the orbital angular momentum
of light \cite[e.g.,][and references therein]{Barnett2022}, particularly
with regard to how the additional unit of momentum contributed by it 
needs to be taken into account in considering the parity and angular 
momentum conservation laws of atom-photon interactions. We leave to
future work the study of how the formalism presented here could be
directly adapted or expanded to the treatment of those interactions 
where both photon spin and orbital angular momentum must play a role.


\begin{acknowledgements}
The authors are deeply indebted with the anonymous referee, who has
painstakingly checked many if not all of the calculations, spotted
several typos, and urged us to clarify various obscure points, 
significantly improving the overall presentation and the potential impact 
of this work. Thank you!

This material is based upon work supported by the National
Center for Atmospheric Research, which is a major facility
sponsored by the National Science Foundation under Cooperative Agreement
No.~1852977.
\end{acknowledgements}

\appendix

\section{Spherical bases}
\label{sec:basis}

Following the definitions of \citetalias{VMK88}, the spherical representation of a vector
$\bm{v}=\sum_q v_q\,\bm{e}_q$, $q=-1,0,1$, adopted by \citetalias{LL04} corresponds to using 
\emph{covariant} coordinates $v_q$ and \emph{contravariant} basis vectors
$\bm{e}_q$, with $\bm{e}_0\equiv\bm{e}_z$. We then have
\begin{displaymath}
\bm{v}=\sum_q v_q\,\bm{e}_q
=\sum_q (-1)^q\,v_q\,\bm{e}^\ast_{-q}
=\sum_q (-1)^q\,v_{-q}\,\bm{e}^\ast_q\;.
\end{displaymath}
Given a different spherical basis $\{\bm{e}_\alpha\}$, $\alpha=-1,0,1$,
we have (see \citetalias{LL04}, eq.~[5.116])
\begin{equation} \label{eq:dotprod}
\bm{v}\cdot\bm{e}_\alpha
=\sum_q (-1)^q\,v_{-q} \left(\bm{e}^\ast_q\cdot\bm{e}_\alpha\right)
=\sum_q (-1)^q\,v_{-q}\,D^1_{\alpha q}(\hat R)\;,
\end{equation}
where $D^K_{QQ'}(\hat{R})$ are rotation matrices 
(\citetalias{VMK88}, \S4.1), and $\hat R$ is the rotation operator 
that transforms the new basis 
$\{\bm{e}_\alpha\}$ into the old basis $\{\bm{e}_q\}$ (or the
vector coordinates from the old basis to the new basis). Thus, if
$(\varphi,\vartheta,\gamma)$ are the Euler angles of the rotation that
carries the old basis into the new one, 
$\hat{R}=\hat{R}_z(-\gamma)\hat{R}_y(-\vartheta)\hat{R}_z(-\varphi) 
	\equiv(-\gamma,-\vartheta,-\varphi)$.

\section{Stokes parameters}
\label{sec:Stokes}

The connection between the second quantization of the radiation field
and the classical coherency matrix (e.g., \citealt{MW95}) is
provided by the following trace operator involving the creation and 
annihilation operators of the various radiation modes 
(see \citetalias{LL04}, Chaps.~4 and 6)
\begin{equation}
I_{\alpha\beta}(\omega,\uk;t)=\frac{1}{8\pi^3}\,\frac{\hbar\omega^3}{c^2}\,
	\mathrm{Tr}\Bigl\{\rho(t)\,
	a^\dagger(\omega,\uk,\alpha)\,a(\omega,\uk,\beta)
	\Bigr\}\;,
\end{equation}
where $\rho(t)$ is the density matrix of the coupled atom-photon
system in the interaction picture. The time dependence of these
ensemble-average quantities is in the coarse-grained sense \citep{LTH71},
and therefore it can be ignored in the limit of 
statistical equilibrium of the coupled system.

From the coherency matrix, we can introduce irreducible spherical 
radiation tensors 
\begin{equation} \label{eq:Itens}
{\cal I}^{l:K}_Q(\omega,\uk)_\mathrm{E,M}=\sum_{\alpha\beta=\pm1}I_{\alpha\beta}(\omega,\uk)\,
{\cal E}^{l:K}_Q(\beta,\alpha;\uk)_\mathrm{E,M}\;,
\end{equation}
as well as the Stokes parameters (cf.~\citetalias{LL04}, eq.~[5.129])
\begin{eqnarray}
\label{eq:Stk-dir}
S_i(\omega,\uk)
&=&\sum_{\alpha\beta=\pm1}(\bm{\sigma}_i)_{\alpha\beta}\,
I_{\beta\alpha}(\omega,\uk)\;, \\
\label{eq:Stk-inv}
I_{\alpha\beta}(\omega,\uk)
&=&\frac{1}{2}\sum_{i=0}^3 
(\bm{\sigma}_i)_{\alpha\beta}\,S_i(\omega,\uk)\;,
\end{eqnarray}
where $\bm{\sigma}_i$, $i=0,1,2,3$, are the Pauli matrices (see
Sect.~\ref{sec:tensors}).
Using eq.~(\ref{eq:Stk-inv}) into eq.~(\ref{eq:Itens}), and recalling
eq.~(\ref{eq:TlKQ.alt}), we find (cf.~\citetalias{LL04}, eq.~[5.150])
\begin{equation} \label{eq:Itens1}
{\cal I}^{l:K}_Q(\omega,\uk)_\mathrm{E,M}=\sum_{i=0}^3 
{\cal T}^{l:K}_Q(i,\uk)_\mathrm{E,M}\, S_i(\omega,\uk)\;,
\end{equation}
which can be inverted using the orthogonality property 
(\ref{eq:TorthoKQ}),
\begin{equation} \label{eq:Stk2}
S_i(\omega,\uk)=\frac{2}{2l+1}\sum_{KQ}
	{\cal T}^{l:K}_Q(i,\uk)_\mathrm{E,M}^\ast\,{\cal
I}^{l:K}_Q(\omega,\uk)_\mathrm{E,M}\;.
\end{equation}
From eq.~(\ref{eq:Itens1}), we see at once that the tensors 
$J^{l:K}_Q(\omega)$ of eq.~(\ref{eq:irreduc_J}) are simply the angular average 
of the radiation tensors ${\cal I}^{l:K}_Q(\omega,\uk)$, i.e.,
\begin{equation} \label{eq:irreduc_J.alt}
J^{l:K}_{Q}(\omega)_\mathrm{E,M}=\oint\frac{d\uk}{4\pi}\;
{\cal I}^{l:K}_{Q}(i,\uk)_\mathrm{E,M}\;.
\end{equation}

\section{Spherical wave expansion of the interaction Hamiltonian}
\label{sec:spherwave}

In order to develop the operator (\ref{eq:HintQ}) as a series of
multipole orders, we rely on the expansion of the exponential factor 
in terms of spherical harmonics and spherical Bessel functions of 
the 1st kind (e.g., \citetalias{VMK88}, eqs.~[5.17.(14)] 
and [5.16.(10)]),
\begin{equation} \label{eq:phasor0}
\exp(\imag\,\vk\cdot\bm{r})=4\pi\sum_{l=0}^\infty\imag^l\,j_l(kr)\sum_m
	Y^l_m(\uk)^\ast Y^l_m(\bm{\hat r})\;,
\end{equation}
and on the spherical-basis representation of a 
vector,
%
\begin{equation} \label{eq:ek0}
\ek{\beta}
=\sum_q D^1_{\beta q}(\hat{R})\,\bm{e}_q
=\sum_q (-1)^q\,D^1_{\beta\,{-}q}(\hat{R})\,\bm{e}_q^\ast\;,
\end{equation}
where $\hat{R}\equiv(-\gamma,-\vartheta,-\varphi)$ is the 
rotation operator that transforms the \emph{helicity} basis 
$\{\ek{\beta}\}$, with $\ek{0}=\uk$, into the spherical 
basis $\{\bm{e}_q\}$, $q=0,\pm1$
(see App.~\ref{sec:basis}, and after
eqs.~[\ref{eq:qmat1.summary}] and 
[\ref{eq:mmat1.summary}]).\footnote{Since we 
adopt a contravariant 
spherical basis $\{\bm{e}_q\}$, the helicity basis 
defined by eq.~(\ref{eq:ek0}) is also contravariant
(cf.\ \citetalias{VMK88}, eqs.~[1.1.(53)]).
Note also that our helicity basis undergoes an additional rotation
by $\gamma$ with respect to the one defined in \citetalias{VMK88};
cf.~their Fig.~1.2 with Fig.~\ref{fig:scatgeom} in this
paper.\label{fn:helicity}}
Therefore,
\begin{equation} \label{eq:phasor}
\exp(\imag\,\vk\cdot\bm{r})\,\ek{\beta}
=4\pi\sum_{l=0}^\infty\imag^l\,j_l(kr)
	\sum_{mq} (-1)^q\,
	Y^l_m(\uk)^\ast Y^l_m(\bm{\hat r})\,
	D^1_{\beta\,{-}q}(\hat R)\,\bm{e}_q^\ast\;.
\qquad (\beta=\pm 1)
\end{equation}

This expression is dealt with more efficiently by using the algebra
of spherical tensors \citepalias{VMK88}. First, we introduce the so-called 
\emph{vector spherical harmonics} (\citetalias{VMK88}, \S7.3). 
\begin{equation} \label{eq:VSH}
\bm{Y}^l_{KQ}\equiv\bigl[\bm{Y}^l\otimes\bm{e}^{\ast 1}\bigr]^K_Q
=\sum_{pq}(lp,1q|KQ)\,Y^l_{p}\,\bm{e}_q^\ast\;,
\end{equation}
where $(a\alpha,b\beta|c\gamma)$ are Clebsch-Gordan (C-G) coefficients
(\citetalias{VMK88}, \S8.1). Using the C-G orthogonality properties, we
have
\begin{displaymath}
Y^l_m(\bm{\hat r})\,\bm{e}_q^\ast
=\sum_{KQ}(lm,1q|KQ)\,
	\bm{Y}^l_{KQ}(\bm{\hat r})\;.
\end{displaymath}
Next, we observe that (\citetalias{VMK88}, eq.~[5.2.7.(37)],
with the symmetry properties of the rotation matrices,
eq.~[4.4.(2)])
\begin{displaymath}
Y^l_m(\uk)^\ast
=(-1)^m\,Y^l_{-m}(\uk)
=\left(\frac{2l+1}{4\pi}\right)\shalf
	(-1)^m\,D^l_{0m}(\gamma,\vartheta,\varphi)
\equiv\left(\frac{2l+1}{4\pi}\right)\shalf
	(-1)^m\,D^l_{0\,{-}m}(\hat{R})\;.
\end{displaymath}
Using these results, we can rewrite eq.~(\ref{eq:phasor}) as
\begin{eqnarray} \label{eq:phasor1}
\exp(\imag\,\vk\cdot\bm{r})\,\ek{\beta}
&=&4\pi\sum_{l=0}^\infty\imag^l\,(-1)^{l+1}\,j_l(kr)
	\left(\frac{2l+1}{4\pi}\right)\shalf
	\sum_{KQ}(-1)^{K+Q}\,\bm{Y}^l_{KQ}(\bm{\hat r})
	\sum_{mq} (lm,1q|K\,{-}Q)\,
	D^l_{0m}(\hat{R})\,D^1_{\beta q}(\hat R) \nonumber \\
&=&4\pi\sum_{l=0}^\infty\imag^l\,(-1)^{l+1}\,j_l(kr)
	\left(\frac{2l+1}{4\pi}\right)\shalf
	\sum_{KQ}(-1)^{K+Q}\,\bm{Y}^l_{KQ}(\bm{\hat r})\,
	(l0,1\beta|K\beta)\,D^K_{\beta\,{-}Q}(\hat R) \nonumber \\
&=&4\pi\sum_{KQ}(-1)^{K+Q}\,D^K_{\beta\,{-}Q}(\hat R) 
	\sum_{l=K-1}^{K+1}\imag^l\,(-1)^{l+1}\,j_l(kr)
	\left(\frac{2l+1}{4\pi}\right)\shalf
	(l0,1\beta|K\beta)\,
	\bm{Y}^l_{KQ}(\bm{\hat r})\;,
\end{eqnarray}
where in the last equivalence we reordered the summations to 
emphasize that the multipole order $l$ is associated with the value 
of $K\ge 1$,\footnote{Since $\beta=\pm 1$ for a transverse field, 
the possibility of $K=0$ is automatically excluded.\label{fn:transverse}} 
and we also 
used the condition of non-nullity of the C-G coefficient 
to limit the summation over $l$ to just the non-vanishing contributions.
For this last summation, explicitly using the fact that 
$\beta=\pm 1$ for a transverse field, we find
\begin{eqnarray} \label{eq:S(K)}
{\cal S}_{KQ}
&\equiv&
	\sum_{l=K-1}^{K+1} \imag^l\,(-1)^{K+l+1}\,j_l(kr)
	\left(\frac{2l+1}{4\pi}\right)\shalf
	(l0,1\beta|K\beta)\,\bm{Y}^l_{KQ}(\bm{\hat r}) \nonumber \\
&=&\imag^{K-1} \left(\frac{2K+1}{8\pi}\right)\shalf
	\Biggl\{\biggl[
	\left(\frac{K+1}{2K+1}\right)\shalf j_{K{-}1}(kr)\,
	\bm{Y}^{K{-}1}_{KQ}(\bm{\hat r})
- 	\left(\frac{K}{2K+1}\right)\shalf j_{K{+}1}(kr)\,
	\bm{Y}^{K{+}1}_{KQ}(\bm{\hat r})
	\biggr] 
	+\imag\beta\,j_{K}(kr)\,\bm{Y}^K_{KQ}(\bm{\hat r})
	\Biggr\} \nonumber \\
&\equiv& {\cal S}_{KQ}^\mathrm{(E)}+{\cal S}_{KQ}^\mathrm{(M)}\;,
\end{eqnarray}
where with the last equivalence we emphasized that the first 
contribution within square brackets is associated with the electric (E)
multipoles of the atomic system, whereas the second one gives rise 
to magnetic (M) multipoles. These two contributions have different parity, 
and they will be considered separately. 

\subsection{Electric-multipole interaction} \label{sec:e-multipole}
In the long-wavelength regime, $kr=(\omega/c)r\ll 1$, the following 
approximations of the spherical Bessel functions hold true,
\begin{equation}  \label{eq:longwave}
j_l(kr)\;\simeq
\kern-18pt\lower9pt\hbox{\tiny{$kr\,{\ll}\,1$}}\;
\frac{(kr)^l}{(2l+1)!!}\;, \qquad
j'_l(kr)\;\simeq
\kern-18pt\lower9pt\hbox{\tiny{$kr\,{\ll}\,1$}}\;
\frac{l}{kr}\,j_l(kr)\;.
\end{equation}
Then, for a given order $l$, the recursion formula
\begin{equation} \label{eq:recursion}
j_{l+1}(z)=\frac{l}{z}\,j_l(z)-j'_l(z)\;,
\end{equation}
shows that $j_{l+1}(z)/j_l(z)\to 0$ in this regime.
This allows us to approximate ${\cal S}_{KQ}^\mathrm{(E)}$ in 
eq.~(\ref{eq:S(K)}) simply as
\begin{eqnarray} \label{eq:S(K)app}
{\cal S}_{KQ}^\mathrm{(E)}
&\simeq&\imag^{K-1}
	\left(\frac{K+1}{8\pi}\right)\shalf
	j_{K-1}(kr)\,
	\bm{Y}^{K{-}1}_{KQ}(\bm{\hat r}) \nonumber \\
&\simeq&\imag^{K-1}
	\left(\frac{K+1}{8\pi}\right)\shalf
	\left(\frac{2K+1}{K}\right)\shalf
	\frac{1}{k}\,
	\mathbf{grad}\bigl\{
	j_K(kr)\,
	Y^K_Q(\bm{\hat r})
	\bigr\}\;,
\end{eqnarray}
where 
in the second line we used eqs.~[7.3.(57-58)] of \citetalias{VMK88}, 
taking again into account the long-wavelength approximation.
%
%
Thus, the electric-multipole contribution to eq.~(\ref{eq:phasor1}) 
for $\beta=\pm 1$ is given by
\begin{equation} \label{eq:phasor3}
\Bigl[\exp(\imag\,\vk\cdot\bm{r})\,\ek{\beta}
\Bigr]_\mathrm{E}
\simeq (2\pi)\shalf 
	\frac{1}{k} \sum_{KQ}
	\imag^{K-1}
	\left[\frac{(K+1)(2K+1)}{K}\right]\shalf
	(-1)^{Q}\,D^K_{\beta\,{-}Q}(\hat R)\;
	\mathbf{grad}\bigl\{
	j_{K}(kr)\,
	Y^K_Q(\bm{\hat r})
	\bigr\}\;,
\end{equation}
%
%
%
%
and the corresponding contribution to the interaction operator (\ref{eq:HintQ}) is
\begin{equation} \label{eq:HintQ1}
Q(\omega,\uk,\beta)_\mathrm{E}
\simeq (2\pi)\shalf\,d_\omega\,
	\frac{1}{k} \sum_{KQ}
	\imag^{K-1}
	\left[\frac{(K+1)(2K+1)}{K}\right]\shalf
	(-1)^{Q}\,D^K_{\beta\,{-}Q}(\hat R)\,
	\sum_i \mathbf{grad}_i\bigl\{
	j_{K}(kr_i)\,
	Y^K_Q(\bm{\hat r}_i)
	\bigr\}\cdot\bm{p}_i\;.
\end{equation}

In order to further develop eq.~(\ref{eq:HintQ1}), 
we introduce the quantum-mechanical operators
\begin{displaymath}
\bm{p}=-\imag \hbar\,\mathbf{grad}\;,\qquad
p^2=-\hbar^2\,\nabla^2\;,
\end{displaymath}
together with the following vector-analysis identity,
\begin{equation}
2\,\mathbf{grad}\,{\cal X}\cdot\mathbf{grad}\,{\cal Y}
	=\nabla^2\bigl({\cal X}\,{\cal Y}\bigr)-
	{\cal X}\,\nabla^2 {\cal Y}-{\cal Y}\,\nabla^2 {\cal X}\;,
\end{equation}
where $\cal X$ and $\cal Y$ are any two functions of $\bm{r}$.
Letting ${\cal X}=F^K_Q\equiv j_{K}(kr)\,Y^K_Q(\bm{\hat r})$, we have
\begin{eqnarray*}
\imag\,\frac{\hbar}{m_{\rm e}}\,
	\mathbf{grad}\,F^K_Q\cdot\bm{p}\,{\cal Y}
&=&\frac{\hbar^2}{2m_{\rm e}}\,\nabla^2\bigl(F^K_Q\,{\cal Y}\bigr)
	-F^K_Q\,\frac{\hbar^2}{2m_{\rm e}}\,\nabla^2 {\cal Y}
	-{\cal Y}\,\frac{\hbar^2}{2m_{\rm e}}\,\nabla^2 F^K_Q \\
&\simeq&\frac{\hbar^2}{2m_{\rm e}}\,\nabla^2\bigl(F^K_Q\,{\cal Y}\bigr)
	-F^K_Q\,\frac{\hbar^2}{2m_{\rm e}}\,\nabla^2 {\cal Y} 
\simeq -\bigl[\mathscr{H}_A,F^K_Q\bigr]\,{\cal Y}\;,
\end{eqnarray*}
where, in the second line, we considered that $F^K_Q$ is harmonic 
in the long-wavelength approximation ($\nabla^2 F^K_Q\simeq 0$), and 
also that $F^K_Q$ \emph{approximately} commutes with the non-kinetic 
part of the atomic Hamiltonian.\footnote{This property 
is rigorously verified for the electrostatic potential, whereas it does 
not hold in the case of the (typically much smaller) spin-orbit
correction, since the spherical harmonics are not eigenfunctions of the
$\bm{L}\cdot\bm{S}$ operator (see also discussion in \citetalias{SM68}, 
\S10.7).} Thus,
\begin{equation}
\mathbf{grad}\bigl\{
	j_{K}(kr)\,
	Y^K_Q(\bm{\hat r})
	\bigr\}\cdot\bm{p}\,
\simeq\imag\,\frac{m_{\rm e}}{\hbar}\,
	\bigl[\mathscr{H}_A,j_{K}(kr)\,
	Y^K_Q(\bm{\hat r})\bigr]\;,
\end{equation}
and after substitution in eq.~(\ref{eq:HintQ1}) and some dimensional
manipulation,
\begin{eqnarray} \label{eq:HintQ2}
Q(\omega,\uk,\beta)_\mathrm{E}
&\simeq& (2\pi)\shalf\,c_\omega\,\frac{1}{k} \sum_{KQ}
	\imag^{K}
	\left[\frac{(K+1)(2K+1)}{K}\right]\shalf
	(-1)^{Q}\,D^K_{\beta\,{-}Q}(\hat R)\, 
	\frac{1}{\hbar}\,
	\bigl[\mathscr{H}_A,\sum_i e\,j_{K}(kr_i)\,
	Y^K_Q(\bm{\hat r}_i)\bigr]\;.
\end{eqnarray}

Ultimately, we are interested in evaluating the matrix element of 
(\ref{eq:HintQ2}) between two atomic states $\ket{m}$ and $\ket{n}$. Using 
again the long-wavelength approximation (\ref{eq:longwave}), we find
\begin{eqnarray} \label{eq:qmat1}
\bigl[Q(\omega,\uk,\beta)_\mathrm{E}\bigr]_{mn}
&\simeq& (2\pi)\shalf\,c_\omega\,\frac{\omega_{mn}}{k} \sum_{KQ}
	\imag^{K}
	\left[\frac{(K+1)(2K+1)}{K}\right]\shalf
	(-1)^{Q}\,D^K_{\beta\,{-}Q}(\hat R)\, 
	\bra{m} \sum_i e\,\frac{(kr_i)^K}{(2K+1)!!}\,
	Y^K_Q(\bm{\hat r}_i)\ket{n} \nonumber \\
&=&\imag\,c_\omega\,\omega_{mn} \sum_{KQ}
	\frac{\imag^{K-1}\,k^{K-1}}{(2K-1)!!}
	\left(\frac{K+1}{2K}\right)\shalf
	(-1)^{Q}\,D^K_{\beta\,{-}Q}(\hat R)\, 
	\bra{m}\!\left(\frac{4\pi}{2K+1}\right)\shalf
	e \sum_i r_i^K Y^K_Q(\bm{\hat r}_i)\ket{n} \nonumber \\
&=&-\imag\,c_\omega\,\omega_{mn} \sum_{K}
	\frac{\imag^{K-1}\,k^{K-1}}{(2K-1)!!}
	\left(\frac{K+1}{2K}\right)\shalf
	\sum_{Q}(-1)^{Q}\,(\mathscr{Q}^K_{-Q})_{mn}\,D^K_{\beta Q}(\hat R)\;,
\qquad (\beta=\pm 1)
\end{eqnarray}
where in the last equivalence we introduced the
\emph{electric-multipole} operator of order $K$,
\begin{equation} \label{eq:Emultipole}
\pmb{\mathscr{Q}}^K\equiv-\biggl(\frac{4\pi}{2K+1}\biggr)\shalf
	e \sum_i r_i^K \bm{Y}^K(\bm{\hat r}_i)\;.
\end{equation}
%
%
In the special case of electric-dipole (E1) transitions ($K=1$), 
eq.~(\ref{eq:qmat1}) gives at once
\begin{equation} \label{eq:E1}
\bigl[Q(\omega,\uk,\beta)_{\rm E1}\bigr]_{mn}
=-\imag\,
	c_\omega\,\omega_{mn}
	\sum_Q (-1)^Q\,(\mathscr{Q}^1_{-Q})_{mn}\,
	D^1_{\beta Q}(\hat{R})\;.
\end{equation}
%
%
For electric quadrupole (E2) transitions ($K=2$), we find instead,
\begin{eqnarray} \label{eq:E2}
\bigl[Q(\omega,\uk,\beta)_{\rm E2}\bigr]_{mn}
&=&\frac{1}{2\sqrt3}\,
	c_\omega\,\omega_{mn}\,k
	\sum_Q (-1)^Q\,(\mathscr{Q}^2_{-Q})_{mn}\,
	D^2_{\beta Q}(\hat{R}) \nonumber \\
&\simeq&\frac{1}{2\sqrt3}\,
	c_\omega\,\frac{\omega_{mn}^2}{c}
	\sum_Q (-1)^Q\,(\mathscr{Q}^2_{-Q})_{mn}\,
	D^2_{\beta Q}(\hat{R})\;,
\end{eqnarray}
where in the last approximation we considered that
$kc=\omega\simeq\omega_{mn}$ near the transition resonance.

%

\subsection{Magnetic-multipole interaction} \label{sec:m-multipole}
In a similar way, we can evaluate the contribution of magnetic multipoles 
to eq.~(\ref{eq:phasor1}). This time, from eq.~(\ref{eq:S(K)}), we have 
\begin{eqnarray} \label{eq:S(K)app_M}
{\cal S}_{KQ}^\mathrm{(M)}
&=&\imag^K
	\left(\frac{2K+1}{8\pi}\right)\shalf
	\beta\,
	j_K(kr)\,
	\bm{Y}^K_{KQ}(\bm{\hat r}) \nonumber \\
&=&\imag^K
	\left(\frac{2K+1}{8\pi}\right)\shalf
	\frac{\beta}{\sqrt{K(K+1)}}\,
	\bm{L}\bigl\{j_K(kr)\,Y^K_Q(\bm{\hat r})\bigr\}\;, \nonumber \\
&=&\imag^{K-1}
	\left(\frac{2K+1}{8\pi}\right)\shalf
	\frac{\beta}{\sqrt{K(K+1)}}\,
	\bigl(\bm{r}\times\mathbf{grad}\bigr)\bigl\{j_K(kr)\,Y^K_Q(\bm{\hat r})\bigr\}\;,
\end{eqnarray}
where we explicitly introduced the quantum operator $\bm{L}$ of the orbital 
angular momentum of the electron, $\hbar\bm{L}=\bm{r}\times\bm{p}
=-\imag\,\hbar\bigl(\bm{r}\times\mathbf{grad}\bigr)$, and we used the 
properties of the vector 
spherical harmonics, eqs.~[7.3.1.(6,9)] of \citetalias{VMK88}. This
leads to the following expression for the magnetic-multipole
contribution to the interaction operator (\ref{eq:HintQ}) for 
$\beta=\pm 1$,
\begin{eqnarray} \label{eq:HintQ1M}
Q(\omega,\uk,\beta)_\mathrm{M}
&=& (2\pi)\shalf\,d_\omega\,\beta \sum_{KQ}
	\imag^{K-1}
	\left[\frac{2K+1}{K(K+1)}\right]\shalf
	(-1)^{Q}\,D^K_{\beta\,{-}Q}(\hat R) 
	\sum_i \bigl(\bm{r}_i\times\mathbf{grad}_i\bigr)\bigl\{
	j_{K}(kr_i)\,
	Y^K_Q(\bm{\hat r}_i)
	\bigr\}\cdot\bm{p}_i \nonumber \\
&=&-(2\pi)\shalf\,d_\omega\,\beta \sum_{KQ}
	\imag^{K-1}
	\left[\frac{2K+1}{K(K+1)}\right]\shalf
	(-1)^{Q}\,D^K_{\beta\,{-}Q}(\hat R) 
	\sum_i \mathbf{grad}_i\bigl\{
	j_{K}(kr_i)\,
	Y^K_Q(\bm{\hat r}_i)
	\bigr\}\cdot\hbar\bm{L}_i\;,
\end{eqnarray}
where in the second line we used the vector-analysis relation
\begin{displaymath}
\bigl(\bm{r}\times\mathbf{grad}\bigr)
F^K_Q\cdot\bm{p}
=-\mathbf{grad}\,F^K_Q\cdot\bigl(\bm{r}\times\bm{p}\bigr)
=-\mathbf{grad}\,F^K_Q\cdot\hbar\bm{L}\;.
\end{displaymath}
In particular, using eq.~[7.3.6.(58)] of \citetalias{VMK88}, we have
\begin{equation} \label{eq:L-contrib}
\mathbf{grad}\bigl\{j_{K}(kr)\,Y^K_Q(\bm{\hat r})\bigr\}
=k \left(\frac{K}{2K+1}\right)\shalf j_{K{-}1}(kr)\,
	\bm{Y}^{K{-}1}_{KQ}(\bm{\hat r})
+k \left(\frac{K+1}{2K+1}\right)\shalf j_{K{+}1}(kr)\,
	\bm{Y}^{K{+}1}_{KQ}(\bm{\hat r})\;.
\end{equation}

We must now calculate the contribution to the interaction Hamiltonian
associated with the electron spin that has so far been neglected, but
which is essential for magnetic-multipole transitions (see discussion in 
the second paragraph of Sect.~\ref{sec:genform}). For this, we must evaluate 
$\mathbf{curl}\,\bm{A}$, which is accomplished by considering 
eqs.~(\ref{eq:phasor1}) and (\ref{eq:S(K)}), 
\begin{displaymath}
\exp(\imag\,\vk\cdot\bm{r})\,\ek{\beta}
=4\pi\sum_{KQ}(-1)^Q\,D^K_{\beta\,{-}Q}(\hat R)\,
{\cal S}_{KQ}\;,
\end{displaymath}
and by using the expressions [7.3.6.(60)] of \citetalias{VMK88} to reduce
$\mathbf{curl}\;{\cal S}_{KQ}$ to its final form. Letting
$\bm{F}^l_{KQ}\equiv j_l(kr)\,\bm{Y}^l_{KQ}(\bm{\hat r})$, we have from
eq.~(\ref{eq:S(K)}),
\begin{eqnarray} \label{eq:S-contrib}
\mathbf{curl}\;{\cal S}_{KQ}
&=&\mathbf{curl}\left[\,{\cal S}_{KQ}^\mathrm{(E)}
	+{\cal S}_{KQ}^\mathrm{(M)}\right] \nonumber \\
&=&\imag^{K-1}\left(\frac{2K+1}{8\pi}\right)\shalf
\Biggl\{\left[
	\left(\frac{K+1}{2K+1}\right)\shalf \mathbf{curl}\,\bm{F}^{K{-}1}_{KQ}
- 	\left(\frac{K}{2K+1}\right)\shalf
	\mathbf{curl}\,\bm{F}^{K{+}1}_{KQ}\right]
	+\imag\beta\,\mathbf{curl}\,\bm{F}^K_{KQ}
	\Biggr\} \nonumber \\
&=&k\,\imag^{K-1} \left(\frac{2K+1}{8\pi}\right)\shalf
\Biggl\{
	-\imag\,\frac{K+1}{2K+1}\,\bm{F}^{K}_{KQ}
	-\imag\,\frac{K}{2K+1}\,\bm{F}^{K}_{KQ}
	+\beta\left[\left(\frac{K}{2K+1}\right)\shalf \bm{F}^{K{+}1}_{KQ}
	-\left(\frac{K+1}{2K+1}\right)\shalf
\bm{F}^{K{-}1}_{KQ}\right]
	\Biggr\} \nonumber \\
&=&k\,\imag^{K-1} \left(\frac{2K+1}{8\pi}\right)\shalf
\Biggl\{
	-\imag\,\bm{F}^{K}_{KQ}
	+\beta\left[\left(\frac{K}{2K+1}\right)\shalf \bm{F}^{K{+}1}_{KQ}
	-\left(\frac{K+1}{2K+1}\right)\shalf
\bm{F}^{K{-}1}_{KQ}\right]
	\Biggr\}\;.
\end{eqnarray}
A comparison of multipole orders between eqs.~(\ref{eq:L-contrib}) and 
(\ref{eq:S-contrib}) shows
that for magnetic-multipole transitions we must retain the terms within 
square brackets in the last line, whereas the first addendum 
$-\imag\,\bm{F}^K_{KQ}$ provides a (typically negligible) spin 
correction to the electric-multipole transitions (see discussion in
\citetalias{SM68}, after eq.~[10.(7.10)]). The contributing term is handled by 
noting that
\begin{eqnarray*}
\mathbf{curl}\;{\cal S}_{KQ}^\mathrm{(M)}
&=&\beta\,k\,\imag^{K-1}\left(\frac{2K+1}{8\pi}\right)\shalf
	\left[\left(\frac{K}{2K+1}\right)\shalf \bm{F}^{K{+}1}_{KQ}
	-\left(\frac{K+1}{2K+1}\right)\shalf\bm{F}^{K{-}1}_{KQ}\right] \\
&=&\beta\,k\,\imag^{K-1}\left[\frac{1}{8\pi}\frac{2K+1}{K(K+1)}\right]\shalf
	\left[K\left(\frac{K+1}{2K+1}\right)\shalf \bm{F}^{K{+}1}_{KQ}
	-(K+1)\left(\frac{K}{2K+1}\right)\shalf\bm{F}^{K{-}1}_{KQ}\right] \\
&=&\beta\,k\,\imag^{K-1}\left[\frac{1}{8\pi}\frac{2K+1}{K(K+1)}\right]\shalf
	\left[\sqrt{(K+1)(2K+1)}\,\bm{F}^{K{+}1}_{KQ}
	-\frac{1}{k}\,(K+1)\,\mathbf{grad}\,F^K_Q\right]\;,
\end{eqnarray*}
where in the last equivalence we used eq.~(\ref{eq:L-contrib}) to
eliminate $\bm{F}^{K{-}1}_{KQ}$. After taking the inner product of 
$\mathbf{curl}\;{\cal S}_{KQ}^\mathrm{(M)}$ with $\hbar\bm{S}_i$ and adding it to the
orbital contribution of eq.~(\ref{eq:HintQ1M}), and some simple
transformations, we find for $\beta=\pm 1$,
\begin{eqnarray} \label{eq:HintQ1M+S}
Q(\omega,\uk,\beta)_\mathrm{M}
&=&-(2\pi)\shalf\,d_\omega\,\beta \sum_{KQ}
	\imag^{K-1}
	\left[\frac{(2K+1)(K+1)}{K}\right]\shalf
	(-1)^{Q}\,D^K_{\beta\,{-}Q}(\hat R) \nonumber \\
&&\kern-1cm {}\times \sum_i \left\{
	\mathbf{grad}_i\bigl\{ j_{K}(kr_i)\,Y^K_Q(\bm{\hat r}_i) \bigr\}
	\cdot\left(\frac{\hbar\bm{L}_i}{K+1}+\hbar\bm{S}_i\right) 
	-k \left(\frac{2K+1}{K+1}\right)\shalf
	j_{K{+}1}(kr_i)\,\bm{Y}^{K{+}1}_{KQ}(\bm{\hat r}_i)\cdot\hbar\bm{S}_i \right\}\;.
\end{eqnarray}

Since ultimately we are interested in the long-wavelength approximation,
the second term in the curly brackets of eq.~(\ref{eq:HintQ1M+S}) can be 
neglected with respect to the first term, similarly to the case of the 
electric-multipole transitions (cf.~eq.~[\ref{eq:S(K)app}]). 
Noting also that, in this approximation,
\begin{displaymath}
\mathbf{grad}\bigl\{ j_{K}(kr)\,Y^K_Q(\bm{\hat r}) \bigr\}
\simeq \frac{k^K}{(2K+1)!!}\,
\mathbf{grad}\bigl\{r^K Y^K_Q(\bm{\hat r}) \bigr\}\;,
\end{displaymath}
the matrix element of the magnetic-multipole interaction operator
between two atomic states $\ket{m}$ and $\ket{n}$ becomes
\begin{eqnarray} \label{eq:mmat1}
\bigl[Q(\omega,\uk,\beta)_\mathrm{M}\bigr]_{mn}
&\simeq&-(2\pi)\shalf\,d_\omega\,\beta \sum_{KQ}
	\frac{\imag^{K-1} k^K}{(2K-1)!!}
	\left[\frac{K+1}{K(2K+1)}\right]\shalf
	(-1)^{Q}\,D^K_{\beta\,{-}Q}(\hat R) \nonumber \\
&&\hphantom{-(2\pi)\shalf\,d_\omega\,\beta \sum_{KQ}}\times 
	\bra{m}\sum_i
	\mathbf{grad}_i\bigl\{ r_i^K Y^K_Q(\bm{\hat r}_i) \bigr\}
	\cdot\left(\frac{\hbar\bm{L}_i}{K+1}+\hbar\bm{S}_i\right)\ket{n} \nonumber \\
&\simeq&c_\omega\,\omega_{mn}\,\beta \sum_K
	\frac{\imag^{K-1} k^{K-1}}{(2K-1)!!}
	\left(\frac{K+1}{2K}\right)\shalf
	\sum_Q (-1)^{Q} (\mathscr{M}^K_{-Q})_{mn}\,D^K_{\beta Q}(\hat R)\;,
\qquad (\beta=\pm 1)
\end{eqnarray}
where in the last equivalence we introduced the
\emph{magnetic-multipole} operator of 
order $K$ (cf.~\citetalias{SM68}, eq.~[10.(6.8$a$)]),
\begin{equation} \label{eq:Mmultipole}
\pmb{\mathscr{M}}^K\equiv-\biggl(\frac{4\pi}{2K+1}\biggr)\shalf \mu_0
	\sum_i \mathbf{grad}_i\bigl\{r_i^K \bm{Y}^K(\bm{\hat r}_i)\bigr\}
	\cdot\left(\frac{2\bm{L}_i}{K+1}+2\bm{S}_i\right)\;.
\end{equation}
and $\mu_0=e\hbar/(2m_{\rm e}c)$ is Bohr's magneton. We note how
the expressions (\ref{eq:mmat1}) and (\ref{eq:qmat1}) are formally
identical, the only difference being the presence of an additional 
factor $\imag\beta$ in the magnetic-multipole case.

\subsection{Alternative form of the spherical-wave expansion}
\label{sec:altwave}

\citetalias{VMK88} provides an expression for the operator of eq.~(\ref{eq:phasor})
fully in terms of vector spherical harmonics, which represents the
starting point for the derivation of multipolar transition matrix elements as
presented by \citetalias{SM68}.

In order to derive this expression (\citetalias{VMK88}, eq.~[7.3.(131)]), we consider
again eqs.~(\ref{eq:phasor0}) and (\ref{eq:ek0}), rewriting the latter
in the explicit form
\begin{displaymath}
\ek{\beta}
=\sum_q (\bm{e}_q^\ast\cdot\ek{\beta})\,\bm{e}_q
\equiv\sum_q (\bm{e}_q\cdot\ek{\beta})\,\bm{e}_q^\ast\;,
\end{displaymath}
where for the second equivalence we used
$\bm{e}_q^\ast=(-1)^q\,\bm{e}_{-q}$. Then, recalling the definition
(\ref{eq:VSH}) of the vector spherical harmonics, we find
\begin{eqnarray*}
\exp(\imag\,\vk\cdot\bm{r})\,\ek{\beta}
&=&4\pi\sum_{l=0}^\infty\imag^l\,j_l(kr)
	\sum_{mq} 
	(Y^l_m(\uk)^\ast\,\bm{e}_q)\cdot\ek{\beta}\,
	(Y^l_m(\bm{\hat r})\,\bm{e}_q^\ast) \\
&=&4\pi\sum_{l=0}^\infty\imag^l\,j_l(kr)
	\sum_{KQ,K'Q'} 
	\bm{Y}^l_{KQ}(\uk)^\ast\cdot\ek{\beta}\,
	\bm{Y}^l_{K'Q'}(\bm{\hat r}) 
	\sum_{mq} (lm,1q|KQ)\,(lm,1q|K'Q') \\
&=&4\pi\sum_{l=0}^\infty\imag^l\,j_l(kr)
	\sum_{KQ} 
	\bm{Y}^l_{KQ}(\uk)^\ast\cdot\ek{\beta}\,
	\bm{Y}^l_{KQ}(\bm{\hat r})\;,
\end{eqnarray*}
which is the result given by \citetalias{VMK88}. Comparison of the last equation with
eq.~(\ref{eq:phasor1}) implies
\begin{equation} \label{eq:needtocheck}
\bm{Y}^l_{KQ}(\uk)^\ast\cdot\ek{\beta}
=(-1)^{K+Q+l+1}
	\left(\frac{2l+1}{4\pi}\right)\shalf
	(l0,1\beta|K\beta)\,D^K_{\beta\,{-}Q}(\hat R)\;.
\end{equation}
In order to prove this equivalence, we first recall that 
our helicity basis $\{\ek{\beta}\}$ is contravariant.
We then use eq.~[7.3.(39)] of \citetalias{VMK88},
together with the straightforward generalization of their
eq.~[7.3.(29)] to the case of $\gamma\ne0$ 
(see footnote \ref{fn:helicity}), to write 
\begin{eqnarray*}
\bm{Y}^l_{KQ}(\uk)^\ast\cdot\ek{\beta}
&=&(-1)^{K+Q+l+1}\,
\bm{Y}^l_{K\,{-}Q}(\uk)\cdot\ek{\beta} \\
&=&(-1)^{K+Q+l+1}
	\left(\frac{2l+1}{4\pi}\right)\shalf
	(l0,1\beta|K\beta)\,D^K_{-\beta\,Q}(\gamma,\vartheta,\varphi)\;.
\end{eqnarray*}
Since
$\hat{R}\equiv(-\gamma,-\vartheta,-\varphi)$, use of the 
symmetry properties of the rotation matrices (e.g., \citetalias{VMK88},
eqs.~[4.4.(2)]) gives
\begin{displaymath}
D^K_{-\beta\,Q}(\gamma,\vartheta,\varphi)=D^K_{\beta\,{-}Q}(\hat{R})\;,
\end{displaymath}
which completes the demonstration of (\ref{eq:needtocheck}).

\section{Dyadic form of the electric-quadrupole interaction operator}
\label{sec:dyadic}

The electric-quadrupole contribution to the interaction Hamiltonian can also be derived 
starting from the ``dyadic'' form presented by \citetalias{LL04} (cf.~eq.~[6.38])
and other authors (e.g., \citealt{Sa67}):
\begin{eqnarray} \label{eq:E2dyad}
\bigl[Q(\omega,\uk,\beta)_{\mathrm{E2}}\bigr]_{mn}
&=&\frac{\imag}{2}\,\frac{e}{m_e}\,c_\omega
	\sum_i\bra{m}\Bigl[
	(\vk\cdot\bm{r}_i)(\ek{\beta}\cdot\bm{p}_i) +
	(\vk\cdot\bm{p}_i)(\ek{\beta}\cdot\bm{r}_i)
	\Bigr]\ket{n} \nonumber \\
&=&\frac{\imag}{2}\,\frac{e}{m_e}\,c_\omega
	\sum_r \sum_i\Bigl[
	\bra{m}\vk\cdot\bm{r}_i\pro{r}\ek{\beta}\cdot\bm{p}_i\ket{n}
	+ \bra{m}\vk\cdot\bm{p}_i\pro{r}\ek{\beta}\cdot\bm{r}_i\ket{n}
	\Bigr] \nonumber \\
\noalign{\eject}
&=&\frac{\imag}{2}\,\frac{e}{m_e}\,c_\omega\,k
	\sum_r \sum_i \imag\,\frac{m_e}{\hbar}\,\Bigl[
	\bra{m}\ek{0}\cdot\bm{r}_i\pro{r}\bigl[\mathscr{H}_A,
	\ek{\beta}\cdot\bm{r}_i\bigr]\ket{n}
\nonumber \\
\noalign{\vspace{-6pt}}
&&\hphantom{\frac{\imag}{2}\,\frac{e}{m_e}\,c_\omega\,k
	\sum_r \sum_i\imag\,\frac{m_e}{\hbar}\kern-5pt}
	+ \bra{m}\bigl[\mathscr{H}_A,\ek{0}\cdot\bm{r}_i\bigr]
	\pro{r}\ek{\beta}\cdot\bm{r}_i\ket{n}
	\Bigr] \nonumber \\
&=&-\frac{e}{2}\,c_\omega\,k
	\sum_r \sum_i\Bigl[
	(\omega_{rn}+\omega_{mr})\,\bra{m}\ek{0}\cdot\bm{r}_i\pro{r}
	\ek{\beta}\cdot\bm{r}_i\ket{n} \Bigr] \nonumber \\
&=&-\frac{e}{2}\,c_\omega\,\omega_{mn}\,k \sum_r \sum_i
	\bra{m}\ek{0}\cdot\bm{r}_i\pro{r}\ek{\beta}\cdot\bm{r}_i\ket{n}
\nonumber \\
&=&-\frac{e}{2}\,c_\omega\,\omega_{mn}\,k \sum_i
	\bra{m}(\ek{0}\cdot\bm{r}_i)(\ek{\beta}\cdot\bm{r}_i)\ket{n}\;,
\end{eqnarray}
where we recalled that $\uk=\ek{0}$, and used the well-known
relation between the electron momentum $\bm{p}$ and the commutator 
$\bigl[\mathscr{H}_A,\bm{r}\bigr]$ (e.g., \citealt{Sa67}, eq.~[2.1.24]).
Next, we observe that 
we can rewrite (see \citetalias{VMK88}, eq.~[1.2.1.(13)], and
eq.~[\ref{eq:dotprod}])
\begin{eqnarray*}
\ek{\alpha}\cdot\bm{r}
=\sum_q (-1)^q\,r_{-q}\,(\ek{\alpha})_q 
=\biggl(\frac{4\pi}{3}\biggr)\shalf r
	\sum_q (-1)^q\,Y^1_{-q}(\bm{\hat r})\,D^1_{\alpha q}(\hat{R})\;,
\end{eqnarray*}
through which eq.~(\ref{eq:E2dyad}) becomes
\begin{eqnarray} \label{eq:QE2}
\bigl[Q(\omega,\uk,\beta)_{\mathrm{E2}}\bigr]_{mn}
&=&-\frac{e}{2}\,c_\omega\,\omega_{mn}\,k \sum_i
	\bra{m}(\ek{0}\cdot\bm{r}_i)(\ek{\beta}\cdot\bm{r}_i)\ket{n}
\nonumber \\
&=&-\frac{4\pi}{3}\,\frac{e}{2}\,c_\omega\,\omega_{mn}\,k
	\sum_i \bra{m} r_i^2 \sum_{qq'}(-1)^{q+q'}\,
	Y^1_{-q}(\bm{\hat r}_i)\,Y^1_{-q'}(\bm{\hat r}_i)\,
	D^1_{0 q}(\hat{R})\,D^1_{\beta q'}(\hat{R})
	\ket{n} \nonumber \\
&\equiv&-\frac{2\pi}{3}\,c_\omega\,\omega_{mn}\,k
	\sum_i 
	\bra{m} e\,r_i^2\,{\cal C}_i(\beta)\ket{n}\;,
\end{eqnarray}
where in the last equivalence we introduced the operator
\begin{displaymath}
{\cal C}_i(\beta)=
	\sum_{qq'}(-1)^{q+q'}\,
	Y^1_{q}(\bm{\hat r}_i)\,Y^1_{q'}(\bm{\hat r}_i)\,
	D^1_{0\,{-}q}(\hat{R})\,D^1_{\beta\,{-}q'}(\hat{R})\;,
\end{displaymath}
which we now evaluate. First, we note that (\citetalias{VMK88}, eq.~[3.1.(22)])
\begin{eqnarray*}
Y^1_{q}(\bm{\hat r}_i)\,Y^1_{q'}(\bm{\hat r}_i)
&=&\sum_{KQ}(1q,1q'|KQ)
	\bigl[\bm{Y}^1\otimes
	\bm{Y}^1\bigr]^K_Q(\bm{\hat r}_i) \\
&=&\sum_{KQ}(-1)^K\,(1\,{-}q,1\,{-}q'|K\,{-}Q)
	\bigl[\bm{Y}^1\otimes
	\bm{Y}^1\bigr]^K_Q(\bm{\hat r}_i)\;,
\end{eqnarray*}
hence
\begin{eqnarray*}
{\cal C}_i(\beta)
&=&\sum_{KQ} (-1)^{K+Q}\,
	\bigl[\bm{Y}^1\otimes
	\bm{Y}^1\bigr]^K_Q(\bm{\hat r}_i)
	\sum_{qq'} (1q,1q'|K\,{-}Q)\,
	D^1_{0 q}(\hat{R})\,D^1_{\beta q'}(\hat{R}) \\
&=&\sum_{KQ} (-1)^{K+Q}\,
	\bigl[\bm{Y}^1\otimes
	\bm{Y}^1\bigr]^K_Q(\bm{\hat r}_i)\,
	(10,1\beta|K\beta)\,
	D^K_{\beta\,{-}Q}(\hat{R})\;.
\end{eqnarray*}
Because of the C-G coefficient, the condition $\beta=\pm1$ 
for the polarization of a transverse field excludes the possibility 
$K=0$ (see footnote \ref{fn:transverse}). In addition, by construction, the tensor 
$\bigl[\bm{Y}^1\otimes \bm{Y}^1\bigr]^K_Q(\bm{\hat r}_i)$
vanishes identical for $K=1$, being the cross-product of $\bm{Y}^1$ with
itself (e.g., \citetalias{VMK88}, eq.~[3.1.(29)]). 
Thus, the summation in ${\cal C}_i(\beta)$ is restricted to just $K=2$. 
Noting finally that 
$(10,1\beta|2\beta)=\sqrt{(4-\beta^2)/6}=1/\sqrt2$ for $\beta=\pm 1$, 
we find
\begin{displaymath}
{\cal C}_i(\beta)
=\frac{1}{\sqrt2}\sum_{Q} (-1)^{Q}\,
	\bigl[\bm{Y}^1\otimes
	\bm{Y}^1\bigr]^2_{-Q}(\bm{\hat r}_i)\,
	D^2_{\beta Q}(\hat{R})\;,
\end{displaymath}
which takes the form of eq.~(\ref{eq:E2}) if we can identify 
the electric-quadrupole operator $\pmb{\mathscr{Q}}^2$ with 
$\sum_i r_i^2\,\bigl[\bm{Y}^1\otimes
\bm{Y}^1\bigr]^2(\bm{\hat r}_i)$. Indeed, this is a direct
consequence of 
$\bigl[\bm{Y}^1\otimes \bm{Y}^1\bigr]^2_{Q}$ 
being simply proportional to the spherical harmonic 
$Y^2_Q$ via a factor $\sqrt{3/(10\pi)}$
(e.g., \citetalias{VMK88}, eq.~[5.6.2.(14)], and Table~8.11). 
Taking this into account, we can rewrite the last equation as
\begin{displaymath}
{\cal C}_i(\beta)
=\frac{1}{2}\biggl(\frac{3}{5\pi}\biggr)\shalf
	\sum_{Q} (-1)^{Q}\,
	Y^2_{-Q}(\bm{\hat r}_i)\,
	D^2_{\beta Q}(\hat{R})\;,
\end{displaymath}
and after substitution into eq.~(\ref{eq:QE2}),
\begin{eqnarray}
\bigl[Q(\omega,\uk,\beta)_{\mathrm{E2}}\bigr]_{mn}
&=&-\frac{1}{2\sqrt3}\,
	c_\omega\,\omega_{mn}\,k
	\sum_{Q} (-1)^{Q}\,
	\bra{m}\,
	e\,\biggl(\frac{4\pi}{5}\biggr)\shalf
	\sum_i r_i^2\,
	Y^2_{-Q}(\bm{\hat r}_i) \ket{n}\,
	D^2_{\beta Q}(\hat{R}) \nonumber \\
&=&\frac{1}{2\sqrt3}\,
	c_\omega\,\omega_{mn}\,k
	\sum_{Q} (-1)^{Q}\,
	(\mathscr{Q}^2_{-Q})_{mn}\,
	D^2_{\beta Q}(\hat{R})\;,
\end{eqnarray}
in agreement with eq.~(\ref{eq:E2}).

\bibliographystyle{aasjournal}
\bibliography{E2_ApJ}

\end{document}